\documentclass[times]{aastex631}
\usepackage{amsmath}	
\newcommand{\be}{\begin{eqnarray}}
\newcommand{\ee}{\end{eqnarray}}
\newcommand{\f}{\Tilde{f}}

\usepackage{multirow}
\usepackage{soul}
\usepackage{enumitem}
\usepackage{xcolor}

\graphicspath{{figs/}}

\begin{document}

\title{What can we learn about Reionization astrophysical parameters using Gaussian Process Regression?}

\author[0000-0002-2701-5654]{Purba Mukherjee}
\affiliation{Centre for Theoretical Physics, Jamia Millia Islamia \\
New Delhi-110025, India}
\email{purba16@gmail.com}

\author[0009-0008-1111-646X]{Antara Dey}
\affiliation{Physics and Applied Mathematics Unit, Indian Statistical Institute \\
203, B.T. Road, Kolkata 700 108, India}
\affiliation{Institute of Physics, Bhubaneswar, Sachivalaya Marg, Odisha 751005, India}
\email{antaraaddey@gmail.com} 

\author[0000-0003-4136-329X]{Supratik Pal}
\affiliation{Physics and Applied Mathematics Unit, Indian Statistical Institute \\
203, B.T. Road, Kolkata 700 108, India}
\email{supratik@isical.ac.in}




\begin{abstract}
Reionization is one of the least understood processes in the evolution history of the Universe, mostly because of the numerous astrophysical processes occurring simultaneously about which we do not have a very clear idea so far. In this article, we use the Gaussian Process Regression (GPR) method to learn the reionization history and infer the astrophysical parameters. We reconstruct the UV luminosity density function using the HFF and early JWST data. From the reconstructed history of reionization, the global differential brightness temperature fluctuation during this epoch has been computed. We perform MCMC analysis of the global 21-cm signal using the instrumental specifications of SARAS, in combination with  Lyman-$\alpha$ ionization fraction data, Planck optical depth measurements and UV luminosity data.  Our analysis reveals that GPR can help infer the astrophysical parameters in a model-agnostic way than conventional methods. Additionally, we analyze the 21-cm power spectrum using the reconstructed history of reionization and demonstrate how the future 21-cm mission SKA, in combination with Planck and Lyman-$\alpha$ forest data, improves the bounds on the reionization astrophysical parameters by doing a joint MCMC analysis for the astrophysical parameters plus 6 cosmological parameters for $\Lambda$CDM model. The results make the GPR-based reconstruction technique a robust learning process and the inferences on the astrophysical parameters obtained therefrom are quite reliable that can be used for future analysis.
\end{abstract}




\section{Introduction}

The epoch of reionization (EoR) represents a crucial period in the evolution history of the Universe, marking the transition from a neutral intergalactic medium (IGM) to one that is fully ionized. This phase, occurring approximately between redshifts $z \approx 6$ and $z \approx 15$, hugely altered the thermal and ionization state of the Universe, setting the stage for the formation and evolution of large-scale cosmic structures \citep[]{Barkana:2000fd, Choudhury:2004vs, Furlanetto:2006jb, Pritchard:2011xb, Kuhlen:2012vy}.  Reionization is primarily driven by the emergence of the first luminous sources, including Population II stars, galaxies, and quasars, which emitted copious amounts of ultraviolet (UV) photons capable of ionizing intergalactic hydrogen. The efficiency of these sources in producing ionizing photons, the fraction of photons that escape into the IGM, and the clumping of the IGM \citep[]{1999ApJ...514..648M, 2009CSci...97..841C, Robertson:2013bq, Bouwens_2017} all play significant roles in shaping the reionization history. These factors collectively influence the reionization power spectrum as well as the global 21-cm signal, both of which serve as key observational probes of this epoch \citep[]{Pritchard:2011xb, Furlanetto:2006jb}. However, despite significant advancements, a complete understanding of the processes that drove reionization and their implications for cosmic evolution remains one of the foremost challenges in contemporary astrophysics.

Along with the lack of sufficient observational data from this epoch, one of the major hurdles towards this direction is a set of astrophysical parameters that govern the epoch of reionization and hence directly impact our understanding of the reionization history and the interpretation of observational data. It is quite a challenging task to constrain these parameters due to their complex interplay and limited observational data, which directly reflects on the (lack of) understanding of the physics of reionization. For instance, the clumping factor \(C_{\rm HII}\), which accounts for the IGM's inhomogeneity, is poorly constrained by observations, with recent simulations suggesting a wide range from 1 to 6 \citep[]{Iliev:2005sz, Pawlik:2008mr, Finlator:2012gr, Schroeder:2012uy}. Secondly, the number of photons entering the IGM depends on the production rate of Lyman continuum (LyC) photons by stars in galaxies, measured by the ionization efficiency $\xi_{\rm ion}$, a parameter that counts ionizing photons per unit UV luminosity. Another astrophysical parameter is the escape fraction \( f_{\text{esc}} \), which is a measure of the fraction of photons entering the IGM thereby ionizing it,  is poorly constrained due to the difficulty in observing the LyC photons beyond \( z \sim 2 - 4.5 \) \citep[]{Inoue:2014zna, Robertson:2021ljt}. Earlier studies suggested a low $\xi_{\rm ion}$ and \( f_{\text{esc}} \approx 0.2 \), fitting more or less well with Cosmic Microwave Background (CMB) data from Planck and Hubble Frontier Fields (HFF) data \citep[]{Robertson:2013bq}. In contrast, the latest James Webb Space Telescope (JWST) findings indicate a higher $\xi_{\rm ion}$, especially at \( z > 9 \), highlighting the degeneracy between $\xi_{\text{ion}}$ and \( f_{\text{esc}} \) \citep[]{Munoz:2024fas, Simmonds_2024, Atek_nature}. Recent studies by \citet{Kulkarni_2019, Finkelstein_2019, Cain_2021, Katz:2022usl} have also investigated any possible redshift evolution of $f_{\text{esc}}$. However, \citet{Mitra:2023yyv} suggest that a constant value of \( f_{\text{esc}} \) between 0.06 and 0.1 for \( z \geq 6 \) is allowed by existing observational data. Thus, within the $\Lambda$CDM framework, magnitude-averaged product of $\xi_{\rm ion}$ and $f_{\rm esc}$ can vary widely depending on the model of reionization, raising significant questions about the validity of cosmological models inferred from reionization data \citep[]{Hazra:2019wdn, Paoletti:2021gzr, Chatterjee:2021ygm, Dey:2022ini, Dey:2023sxx, Paoletti:2024lji}, which in turn reflects on the inference drawn about the reionization history as a whole. 

In light of these widespread uncertainties related to the proper estimation of astrophysical parameters from direct approaches, searches for possible alternative tools that may help in having a somewhat better idea about them from the existing data alone, are natural questions the community has started to ask of late. One such interesting tool is the use of Machine Learning (ML) techniques \citep[]{Rasmussen2006Gaussian, Takezawa, Harezlak}, that can significantly enhance our understanding of the epoch of reionization by developing flexible, data-driven models to analyze observational data. Unlike traditional methods, which depend on the assumption of predefined models that may overlook key details, ML techniques can uncover hidden patterns and relationships in complex data sets. These advanced inference methods can provide a better understanding of the astrophysical parameters, accounting for their variations across different redshifts \citep[]{Krishak:2021fxp, Mitra:2023yyv}, leading to unbiased, model-independent reconstructions of the reionization history and a relatively deeper insight into the underlying physics. In the present article, we seek to elucidate the factors driving reionization and contribute to mapping the Universe's reionization history. As it will turn out, this will stem from a somewhat better hold on the astrophysical parameters through an ML-based learning process. 

Further, one of the most exciting probes of the EoR is the spin-flip line of neutral hydrogen, with a rest frame wavelength of 21-cm. Advancements in current instruments and upcoming missions are expected to revolutionize  21-cm cosmology from observational point of view, providing highly significant measurements of both the power spectrum and the global 21-cm brightness temperature signal \citep[]{Pritchard:2011xb}. Currently, the Shaped Antenna measurement of the background RAdio Spectrum (SARAS) \citep[]{2013ExA....36..319P} aims to measure the global sky-averaged 21-cm signal from the cosmic dawn and the EoR using a shaped antenna to capture the redshifted 21-cm line \citep[]{Singh:2017cnp}. In the coming decade, experiments like the Square Kilometre Array (SKA) \citep[]{2015ExA....39..567D}, Hydrogen Epoch of Reionization Array (HERA) \citep[]{HERA:2021noe} and other significant missions \citep[]{2013A&A...556A...2V}, will drive advancements in cosmology by detecting the  21-cm neutral hydrogen signal from the early Universe. By employing 21-cm intensity mapping techniques, SKA will track neutral hydrogen in the Universe, yielding comprehensive insights into the post-reionization and reionization epochs, as well as the cosmic dawn, up to a redshift of 30 \citep[]{Pritchard:2011xb}. Our study explores how the ongoing SARAS  and next-generation SKA, with their innovative approaches, will enhance our understanding of reionization by reconstructing the reionization history and refining the bounds on the associated astrophysical parameters. 

In this work, we intend to enhance the understanding of reionization by examining the effects of various astrophysical parameters on its timeline. We begin by reviewing the theoretical framework that describes the ionization state of the IGM, highlighting significant astrophysical parameters and their connection to observables {in sections \ref{sec:astro} and \ref{sec:theory}. Section \ref{sec:data} presents the observational data relevant to our work. In section \ref{sec:method},} we employ an ML algorithm, Gaussian Process Regression\footnote{\url{https://gaussianprocess.org/gpml/}} (GPR) to perform a Bayesian, non-parametric, model-independent reconstruction \citep[]{Rasmussen2006Gaussian, Takezawa, Harezlak} of UV luminosity density $\log_{10} \rho_{\text{\rm UV}}$ as a function of redshift, using the current observations from Hubble Frontier Fields (HFF) \citep[]{Lotz:2016owk, Schenker:2012vs, Ellis:2012bh, McLure:2012fk, McLeod_2016, Oesch_2018} compiled by \citep[]{Bouwens:2014fua, Bouwens_2017, Bowens_2021}, early James Webb Space Telescope (JWST) \citep[]{Harikane_2023}, and Subaru HSC's Great Optically Luminous Dropout Research data \citep[]{goldrush}. {GPR, is a non-parametric regression technique that estimates functions based on a Gaussian prior distribution over possible predictions. It provides a flexible, data-driven approach to reconstructing smooth functions from irregularly spaced and heterogeneous data, making it particularly suited for our study. More details on the algorithm, including its historical perspectives and cosmological context, have been discussed previously by some of the authors of the present manuscript in \citet{Mukherjee:2023lqr}. A joint evaluation of both the GPR kernel hyperparameters and mean function parameters, allows one to balance the physics-based model with the ML model. Thus,} our approach allows for flexible modeling of the evolution of ionizing sources without imposing restrictive parametric forms \citep[]{Ishigaki_2015, Ishigaki_2018, Adak:2024urf}. 

{After reconstructing the $\log_{10} \rho_{\text{\rm UV}}$ profile, we then adopt GPR to interpolate between four binned redshifts within $4<z<12$, to obtain a smooth and model-independent representation of the reionization history as a function of redshift. For this exercise,} we consider the neutral hydrogen fraction measurements \citep[]{Greig:2016vpu, Davies:2018yfp, 2006PASJ...58..485T, 2008MNRAS.388.1101M, 2011MNRAS.416L..70B, 2011Natur.474..616M, 2012ApJ...744...83O, 2014ApJ...795...20S, 2014ApJ...794....5T, 2019MNRAS.485.3947M} and optical depth $\tau_{\text{reio}}$ constraints from the Planck 2018 release of CMB observations \citep[]{refId0}, in addition to UV luminosity data sets. We undertake a full Bayesian Markov chain Monte Carlo (MCMC) analysis to explore the role of existing data sets for predicting the observationally favoured bounds on the reionization astrophysical parameters. {The Bayesian framework of GPR provides valuable advantages in this regard, both as a maximum likelihood estimator (MLE) and as an interpolant between redshift bins.}

In what follows, we detail our methodology for analyzing the global 21-cm signal and the reionization power spectrum within the framework of the $\Lambda$CDM cosmological model. For the global 21-cm signal $\Delta T_b$, we generate a mock $\Delta T_b$ vs $z$ data, assuming the instrumental specifications of SARAS \citep[]{2013ExA....36..319P}, considering the best-fit values of these astrophysical parameters obtained using the existing data sets. For the power spectrum analysis, we use the instrumental specifications of SKA to generate a mock data set \citep[]{SKALowv2}, using the Planck 2018 best-fit $\Lambda$CDM model. We make prior modifications in \texttt{CLASS} \citep{Blas_2011} to incorporate the contribution from the GP reconstructed reionization history profile (instead of the inbuilt $\tanh$ reionization model), to compare both the cases and reflect on their outcomes. In the final step, we conduct a comprehensive Bayesian MCMC analysis to investigate how SARAS and upcoming SKA will aid in probing the astrophysical parameters of reionization. {We summarize our findings in section \ref{sec:results}, and make some concluding remarks in section \ref{sec:conclusion}}.

{
\section{Astrophysical Parameters and the role of Machine Learning}
\label{sec:astro}

The ionization equation describes the time evolution of the volume filling factor of ionized hydrogen in the intergalactic medium, $Q_{\rm HII}$, by a first-order ordinary differential equation,
\be
   \frac{\mathrm{d}Q_{\rm HII}}{\mathrm{d}t}= \frac{\dot{n}_{\text{ion}}}{\langle n_{\rm H}\rangle} -\frac {Q_{\rm HII}}{t_{\text{rec}}} \, .
   \label{eq:ioneqn}
\ee
The source term $\dot{n}_{\text{ion}}$ is characterized by the rate of production of ionizing photons, which depends on (i) the UV luminosity density function $\rho_{\rm UV}$, (ii) the efficiency of the source to produce ionizing photons $\xi_{\text{ion}}$, (iii) the fraction of photons that escape into the IGM $f_{\text{esc}}$. It is defined as,
\be 
\dot n_{\text{ion}}=  \rho_{\rm UV}\langle f_{\text{esc}}\xi_{\text{ion}}\rangle \, , \label{sec:source}
\ee
where $\langle f_{\text{esc}}\xi_{\text{ion}}\rangle$ is a magnitude-averaged product. The sink term in the ionization equation accounts for the recombination process in the IGM. The recombination time $t_{\text{rec}}$, given by
\be 
t_{\text{rec}}=\left[ {C_{\rm HII} \, \alpha_B(T)\left(1+\frac{Y_p}{4X_p}\right)\langle n_{\rm H}\rangle (1+z)^3} \right]^{-1}  \, 
\ee
where $n_{\rm H}, n_{\rm He}, n_{\rm HII}$ are the number densities of hydrogen, helium, and ionized hydrogen, is determined by the recombination coefficient $\alpha_B(T)$ and the clumping factor $C_{\rm HII}$. Here, $X_p, Y_p$ are the primordial mass fractions of hydrogen and helium. The $C_{\rm HII}$ accounts for the inhomogeneity of the IGM, and is not very well constrained from observations. The IGM temperature $T$ is fixed at 20,000 K.  

From the solution of the ionization Eq. \eqref{eq:ioneqn}, the Thomson scattering optical depth is defined as,
\be
\tau = \int \frac{c \, (1+z)^2}{H(z)} \, Q_{\rm HII}(z) \, \langle n_{\rm H} \rangle \, \sigma_{\rm T} \left( 1+ \eta \, \frac{Y_p}{4 X_p}\right) \mathrm{d} z \, , \label{eq:tau}
\ee
where helium is assumed to be singly-ionized for $z > 4$ ($\eta$ = 1) and doubly-ionized for $z < 4$ ($\eta$ = 2) \citep{Kuhlen:2012vy}.

{Thus, the astrophysical parameters play a crucial role in driving the reionization history as well as in constraining the underlying cosmological model parameters therefrom. However, in the absence of sufficient information about these astrophysical parameters, either from the 21-cm signal or complementary probes, their range of allowed values remains significantly wide. As a result, it is unlikely to draw reliable conclusions about the cosmological model and the reionization history. So, to have a better understanding of the scenario, the community has to wait for improved data, hopefully leading to more precise values of the astrophysical parameters. Alternatively, one can utilize statistical/numerical tools to predict the nature of reionization history in a non-parametric way by making use of the existing data from complimentary probes like HST and JWST, and infer the astrophysical parameters therefrom. Herein lies the major role of ML techniques as an interesting route to progress even with currently available data sets.

In this work, we employ a particular ML technique- Gaussian Process Regression (GPR)- to infer the astrophysical parameters that would in turn help us have a better understanding of the reionization history as well as a better hold on the cosmological parameters in the reionization era. To this end,} we attempt to first reconstruct the UV luminosity density using GPR in a non-parametric way {with the UV LF data}. Subsequently, we will use the reconstructed values of $\log_{10} \rho_{\text{\rm UV}}$, interpolated at four binned redshifts within $4<z<12$, {in combination with the volume-averaged neutral hydrogen fraction ($Q_{\rm HII}$), and the Thomson optical depth ($\tau_{\rm reio}$) measurements}, to derive a model-independent reionization history profile as a function of redshift and infer the astrophysical parameters $\langle f_{\text{esc}}\xi_{\text{ion}}\rangle$ and $C_{\rm HII}$, respectively. 

\section{21cm-cosmology and observables} \label{sec:theory}

In this section, extend our analysis to the yet-unexplored directions on reionization studies. This is materialized by considering the global 21-cm signal and 21-cm power spectrum, exploring their potential in inferring the reionization physics. 

}

\subsection{Global Brightness Temperature Fluctuation} \label{sec:Tb_21}

The key observable in 21-cm cosmology is the global brightness temperature fluctuation, which is the difference between the spin temperature (related to the neutral hydrogen number densities in different atomic levels) and the background temperature. The total brightness temperature at redshift $z$ is given by the temperature of the background radiation field, with some fraction of it absorbed and re-emitted due to 21-cm hyperfine transitions in neutral hydrogen atoms. The properties of HI in absorption and emission are described by the spin temperature $T_{s}$ and the optical depth $\tau$ \citep[]{Pritchard:2011xb, Furlanetto:2006jb}:
\be
T_{\rm b} = T_s(1-e^{-\tau})+T_{\gamma}e^{-\tau} \, .
\ee
Due to the low probability of a 21-cm transition, the optical depth is typically small. The differential brightness temperature can thus be written as linear in $\tau$:
\be
\label{DTb_tau}
\Delta T_{\rm b} = \frac{T_s-T_{\gamma}}{1+z}\left(1-e^{-\tau}\right) \approx \frac{T_s-T_{\gamma}}{1+z}\tau \, .
\ee

The optical depth produced by a patch of neutral hydrogen at the mean density and with a uniform 21-cm spin temperature $T_{s}$,
\be
\tau=9.0 \times 10^{-3} \left (\dfrac{T_{\text{CMB}}}{T_{s}} \right)\left (\dfrac{\Omega_{\rm b}h}{0.03} \right) \left(\dfrac{\Omega_{\rm m}}{0.3} \right)^{-1/2} \left (\dfrac{1+z}{10}\right)^{1/2} \, .
\ee

The Lyman-$\alpha$ (Ly$\alpha$) and X-ray radiation backgrounds during the epoch of reionization are anticipated to be strong enough to equalize the spin temperature $T_{s}$ with the gas temperature and heat up the cosmic gas well above the CMB temperature \citep[]{Madau:1996cs}. In these circumstances, the observed 21-cm brightness temperature $T_{b}$, in relation to the CMB temperature $T_{\gamma}$, becomes independent of $T_{s}$. Consequently, $T_{b}$ (hereafter measured relative to $T_{\gamma}$) is given by \citep[]{2012MNRAS.424.2551M}
\be
\Delta T_{b}=(T_{s}-T_{\gamma})(1- e^{-\tau}) ~ Q_{\rm HI} \approx T_{21} \left (\dfrac{1+z}{10} \right)^{1/2} Q_{\rm HI} \, , \label{eq:deltaTb}
\ee
where $T_{21} = 9.0 \times 10^{-3} (\Omega_{\rm b}h/0.03) (\Omega_{\rm m}/0.3)^{-1/2} \,  T_{\text{CMB}} = 27.2$ mK and $Q_{\rm HI}$ is the neutral hydrogen fraction. We focus solely on the cosmic mean neutral or ionized fraction and disregard spatial fluctuations in the 21-cm signal caused by density and peculiar velocity variations.

\subsection{Power Spectrum} \label{sec:Pk_21}

The difference between the 21-cm temperature $T_{b}(\mathbf x)$ and the average temperature $\overline T_{b} (z)$ at a given redshift can be calculated at any spatial point and is denoted by $\Delta T_{b} (\bf x)$. Its Fourier transform is indicated as $\Delta T_{b} (\bf x)$. The two-point correlation function of 21-cm temperature fluctuations at redshift $z$ is written as \citep[]{Munoz:2016owz}
\be
{\Delta T_{b} (\mathbf k) \Delta T_{b} (\mathbf k')}\equiv P_{21}(\mathbf k,z)  (2\pi)^3 \delta_D{(\bf k-k')} \, ,
\ee
with
\be
P_{21}(\mathbf k,z)= \left [ {\cal A}(z) + \overline {T}_{b} (z) \mu^2\right]^2 P_{\rm HI}(\mathbf k,z) \, ,
\label{eq:P21}
\ee
here, ${\cal A}(z) = \mathrm{d} T_{21}/\mathrm{d} \delta_b$ indicates a function of $z$, $\mu\equiv k_{||}/k$ is the cosine of the angle between the line-of-sight $k_{||}$ and the total wave vector $k = \vert \mathbf{k} \vert$, and $P_{\rm HI}$ is the power spectrum for the perturbations in neutral hydrogen density, {which is related to matter power spectrum $P_{m}(\mathbf k, z)$ with a bias $b_{\rm HI}$. We have used the bias model \( b_{\rm HI}(z) = \beta_{0} \left( 0.904 + 0.135 (1+z)^{0.198 \beta_{1}} \right) \), where \( \beta_{0} \) and \( \beta_{1} \) are treated as nuisance parameters \citep[]{Sprenger:2018tdb}.}

The spin temperature $T_s$ during the EoR is coupled to the gas temperature through the Wouthuysen-Field effect \citep[]{1952AJ.....57R..31W, Hirata:2005mz}. The star formation heats the gas, which give rise the spin temperature above the CMB temperature $T_\gamma$, making the 21-cm line appear in emission. In this epoch, we can express the factors in Eq. \eqref{eq:P21} as, \citep[]{Munoz:2016owz}

\be
\mathcal A(z) = \overline T_{21} (z) = 27.3\, {\rm mK}\,\times \overline x_{\rm H} \dfrac{T_s - T_{\gamma}}{T_s} \left ( \dfrac{1+z}{10} \right)^{1/2} \, ,
\label{eq:T21bar}
\ee
here we can drop the temperature factor since $T_s\gg T_\gamma$, and  $\overline x_{\rm H}$ represents the mean neutral hydrogen fraction 

Assuming an antenna array with a baseline $D_{\rm base}$ uniformly covered to a fraction $f_{\rm cover} \leq 1$ and an observation time of $t_o$, the instrumental-noise power spectrum in $k$-space is given by \citep[]{Zaldarriaga:2003du, Tegmark:2008au}
\be
P^N_{21} (z) = \dfrac{\pi T_{\rm sys}^2}{t_o f_{\rm cover}^2} \chi^2(z) y_\nu(z) \dfrac{\lambda^2(z)}{D_{\rm base}^2} \, ,
\ee
In this context, $\lambda(z)$ is the 21-cm transition wavelength corresponding to redshift $z$, $y_\nu(z)= 18.5,\sqrt{(1+z)/10}$ Mpc/MHz serves as the conversion function from frequency $\nu$ to $k_{||}$, and the system temperature $T_{\rm sys}$ is predominantly determined by galactic synchrotron emission, characterized as \citep[]{deOliveira-Costa:2008cxd}
\be
T_{\rm sys} = 180 \, {\rm K} \times \left( \dfrac{\nu}{180\,\rm MHz}\right)^{-2.6} \, .
\label{eq:Tsys} 
\ee

Combining all these pieces of information, the observed power spectrum looks like \citep[]{Sprenger:2018tdb, Dey:2023sxx}
\be
     P_{\rm 21,obs}(\mathbf k,\mu,z)= f_{\rm AP}(z) \times f_{\rm res}(k,\mu,z) \times f_{\rm RSD}(\hat{k},\hat{\rm \mu},z) \times P_{\rm 21}(\mathbf k,z) + P^N_{\rm 21} (z) \, ,
\ee
where $P_{\rm 21}(\mathbf{k},z)$ denotes the 21-cm power spectrum. In the above formula, we have applied the flat-sky approximation, which provides a specific definition of the line-of-sight distance vector $\Vec{r}$ and Fourier modes. This approximation breaks the isotropy along the observer's line of sight but retains the symmetry perpendicular to it. The coordinate relations are as follows: $k = |\Vec{k}|$, $\mu = \dfrac{\Vec{k} \cdot \Vec{r}}{k r}$, with the parallel component of the mode being $k_{\rm \parallel} = \mu k$ and the perpendicular component being $k_{\rm \perp} = k \sqrt{1 - \mu^{2}}$.

\section{Data sets \label{sec:data}}

In the process of learning the reionization history, we have two-fold goals: (i) to find out the present constraints from a couple of cosmological data sets that also help in the reconstruction process and (ii) to forecast on the astrophysical parameters along with the cosmological parameters from the 21-cm power spectra. \\

\noindent For current constraints, the data sets used are the following:
\begin{itemize}

\item {The UV luminosity density $\rho_{\rm UV}$ data, derived by integrating the UV LF, $\Phi(M_{\rm UV})$, as $\rho_{\rm UV} = \int_{M_{\rm trunc}}^{-\infty} \Phi(M) \, L(M) \, \mathrm{d}M $, where $L(M)$ represents the luminosity at a given absolute magnitude $M_{\rm UV}$. The integration limit is set at a conservative truncation magnitude $M_{\rm trunc}=-17$, corresponding to the minimum observed halo mass that can host star-forming faint galaxies. The estimation of $\rho_{\rm UV}$ relies on an accurate characterization of the LF profile, which describes the number density of star-forming galaxies at different luminosities, modeled via the \textit{Schechter} function \citep{Schechter:1976iz}. Herein, we use two compilations of UV luminosity from various deep-field surveys,}
\begin{itemize}[left=0pt]
    \item \textbf{UV17(A)}:  The derived UV LF density data \citep[]{Bouwens:2015vha, Bouwens_2017} at $z$ $\sim$ 4-10 from HFF observations\footnote{\url{http://www.stsci.edu/hst/campaigns/frontier-fields/}} \citep[]{2017ApJ...837...97L}, {illustrated in Fig. \ref{fig:rho_uv}(b)}.
    \item \textbf{UV17(B)}: The UV LF data for \( z \sim 2-7 \), derived in \citet{Adak:2024urf} using the HUDF, HFF, and CANDELS fields, compiled by \citep[]{Bouwens:2015vha, Ishigaki_2018}. For \( z \sim 4-7 \), we incorporate data from the Hyper Suprime-Cam (HSC) Subaru Strategic Program (SSP) survey \citep[]{goldrush}. For \( z \sim 8-10 \), we consider the derived LF obtained by \citet{Bouwens:2022gqg, Adak:2024urf} and early JWST data at \( z \sim 9 \) and 12 \citep[]{Harikane_2023}. {For visualization, see Fig. \ref{fig:rho_uv}(c)}.
\end{itemize}

\item \textbf{QHII}: Neutral hydrogen fraction ($Q_{\rm HII}$) measurements {from multiple probes, illustrated in Figs. \ref{fig:qhii_Tb_plots}(a)-(b), providing complementary insights into the patchy nature of reionization across cosmic epochs.}
\begin{itemize}[left=0pt]
    \item Ly$\alpha$ emission from galaxies \citep[]{2012ApJ...744...83O, 2014ApJ...795...20S, 2014ApJ...794....5T, 2019MNRAS.485.3947M}
    \item Damping wing absorption signatures of gamma-ray bursts \citep[]{2006PASJ...58..485T, 2008MNRAS.388.1101M}
    \item Dark gap in quasar spectra, indicating neutral intergalactic medium regions \citep[]{2015MNRAS.447..499M}
    \item Ionized proximity zones near high redshift quasars, to estimate the surrounding ionization state \citep[]{2011Natur.474..616M, 2011MNRAS.416L..70B}
\end{itemize}

\item \textbf{Planck}: The Thomson scattering optical depth constraints $\tau_{\rm reio} = 0.054 \pm 0.007$ from Planck 2018 release of CMB observation \citep[]{refId0}, {as an integrated measure of the reionization history (illustrated via dashed lines with shaded regions in Fig. \ref{fig:mcmc_present}. We adopt the mean and the 1$\sigma$ confidence limit of $\tau_{\rm reio}$ from the Planck TTTEEE+lowE+lensing likelihood as summary statistics. Instead of using the full Planck MCMC chains, we incorporate this constraint as a Gaussian prior on $\tau_{\rm reio}$, ensuring consistency with CMB-derived reionization models.}
\end{itemize}

On the other hand, for future forecasts from 21-cm observables, viz. the global brightness temperature fluctuations and power spectra, we make use of the following data sets:

\begin{itemize}

\item \textbf{SARAS}: Mock data generated from the instrumental specifications of the global 21-cm mission \citep[]{2013ExA....36..319P}, {incorporated alongside the existing observational datasets. For a visualization of the generated mock $\Delta T_b$ vs $z$, one can refer to Fig. \ref{fig:saras}(d)}.

\item \textbf{Planck}: Mock Planck 2018 data  \citep[]{refId0, Sprenger:2018tdb}, generated for forecast analysis, making it suitable for combination with other future experiments using \texttt{MontePython} \citep{Audren:2012wb,Brinckmann:2018cvx}. {It is derived by sampling from the posterior distribution of Planck MCMC chains, ensuring realistic uncertainties and statistical properties, incorporating the theoretical CMB power spectra predictions, instrumental noise, and observational constraints.}

\item \textbf{Ly$\alpha$}: MIKE-HIRES Lyman-$\alpha$ forest dataset \citep[]{Viel:2013fqw}, obtained from QSO spectra measured with the HIRES/KECK and MIKE/Magellan spectrographs, has been used. These observations span redshift bins $z=5.0$ and $z=5.4$, with spectral resolution of 13.6 km/s (HIRES) and 6.7 km/s (MIKE), and are binned into 10 $k$-bins over the range $0.001-0.08$ s/km. We impose a conservative cut on the flux power spectrum, including only measurements with $k>0.005$ s/km to minimize potential systematic uncertainties from continuum fitting at large scales.

\item \textbf{SKA}: Simulated data using the instrumental specifications of SKA-Low \citep[]{SKALowv2}, {generated incorporating noise modeling and conservative foreground mitigation strategies to mimic realistic survey conditions. Our key observable is the 21-cm power spectrum, to forecast the expected 21-cm signal from reionization epochs, enabling constraints on reionization parameters and astrophysical models.}

\end{itemize}

\section{Methodology \label{sec:method}}

\subsection{Reconstructing the UV luminosity density \label{sec:uvlum}}

The evolution of the UV luminosity density with redshift can be obtained by parametric \citep{Yu_2012, Ishigaki_2015, Ishigaki_2018, Adak:2024urf} and non-parametric free-form methods \citep{Hazra:2019wdn, Paoletti:2021gzr} to determine $n_{\text{ion}}$. Recently, a model-independent reconstruction of $\rho_{\rm UV}$ by \citep{Krishak:2021fxp} invalidates the single power-law form \citep{Yu_2012}, as it fails to account for the decline at $z \sim 8$, resulting in an incorrect Thomson scattering optical depth. Therefore, the assumption of a parametric logarithmic double power law \citep{Ishigaki_2015, Ishigaki_2018}, given by 
\be
\rho_{\rm UV}(z)&=&\frac{2\rho_{{\rm UV},z=z_1}}{10^{a(z-z_1)}+10^{b(z-z_1)}} \, , \label{eq:powerlaw2}
\ee
to describe the UV LF profile is a better ansatz, characterised by four distinct parameters, namely - the amplitude ($\rho_{{\rm UV},z=z_1}$), two tilts ($a,b$) and the redshift ($z_1$) at which the tilt in the power changes. 

While parametric methods are useful, the functional form restricts their ability to address the data in several instances [see \citep[]{Yu_2012, Ishigaki_2018, Adak:2024urf}]. So, a more robust approach is a non-parametric reconstruction which attempts to reconstruct the cosmic reionization history directly from the observational data. In this article, we use GPR \citep[]{Rasmussen2006Gaussian, Seikel:2012uu, Shafieloo:2012ht, Mukherjee:2022lkt, Shah:2023rqb, Mukherjee:2023lqr} for a Bayesian, non-parametric reconstruction \citep[]{Takezawa, Harezlak} of luminosity density in a model-independent manner. 

\subsection{Gaussian Process Framework}

A Gaussian Process (GP) is a collection of random variables such that the joint distribution of any finite subset of it is described by a multivariate Gaussian. It is characterized by a mean function $\mu(\mathrm{x})$ and covariance function  $k(\mathrm{x},\mathrm{x'})$, where for a real process $f(\mathrm{x})$, we have $\mu(\mathrm{x}) = \mathbb{E}[f(\mathrm{x})]$, and $k(\mathrm{x},\mathrm{x'}) = \mathbb{E}[(f(\mathrm{x})-\mu(\mathrm{x}))(f(\mathrm{x'})-\mu(\mathrm{x'}))]$. 

For a finite set of training points $\mathrm{x}=\{x_i\}$, a function $f(\mathrm{x})$ evaluated at each $x_i$ can be represented by a random variable with a Gaussian distribution, such that the vector $\mathrm{f}=\{f_i\}$ has a multivariate Gaussian distribution given as $\mathrm{f}\sim \mathcal{N}\left(\mu(\mathrm{x}), C(\mathrm{x},\mathrm{x})\right)$, where $C$ is the covariance matrix characterized by the kernel or covariance function $k(x_i, x_j)$, which gives the covariance between two random variables $x_i$ and $x_j$ respectively. 

For our analysis, we choose the Radial Basis Function (RBF) kernel, represented as $ k(x_i,x_j)= \sigma_f \, \text{exp}\left[- \frac{(x_i-x_j)^2}{2l^2}\right]$, with the correlation length $l$ and amplitude $\sigma_f$. The logarithmic double power law parametrization, given in Eq. \eqref{eq:powerlaw2} is considered a mean function, whose parameters are jointly constrained with the kernel hyperparameters, marginalizing the log-likelihood via a Bayesian MCMC analysis with \texttt{emcee} \citep{Foreman-Mackey:2012any}. {The constraints on the kernel hyperparameters and parameters governing the mean function are shown in Fig. \ref{fig:rho_uv}(a).} 

We undertake this exercise for both UV17(A) and UV17(B) compilations of the luminosity density data as the training set. Although our reconstruction method is somewhat in the same vein of \citet[]{Krishak:2021fxp}, where the mean function is fixed to the best-fit values obtained by \(\chi^2\) minimization, our novelty lies in simultaneously training the parameters governing the GP mean function and the kernel hyperparameters to obtain the predicted UV luminosity density profile, which helps the predictions arise from a symbiotic environment and hence is expected to generate more realistic outcome of the learning process.

\subsection{Learning the reionization history \label{sec:reion}}

On having reconstructed the profile of UV luminosity density in a model-independent way, we now re-define the values of $\log_{10} \rho_{\rm UV}$ in four distinct equidistant nodes in the range $4 <  z < 12$. Our approach is akin to \citet{Gerardi:2019obr}, where this range is selected to fully encompass available UV17(A) and UV17(B) data sets. The values of the UV luminosity density at these four nodes i.e., \( \log_{10} \rho_{1-4} \) are taken as free parameters in MCMC sampling, employing \texttt{emcee} \citep{Foreman-Mackey:2012any}, to learn the reionization history by solving the ionization Eq.  \eqref{eq:ioneqn}  using this model-independent form. At each MCMC step, these points serve as training data for GP regression. Hence, GP reconstruction yields samples of the history of UV luminosity densities, based on the training input configurations.  

For this full Bayesian analysis, we take into account different combinations of the existing data sets, described in Sec. \ref{sec:data}. In the ionization equation, we treat $ \left\langle f_{\text{esc}} \xi_{\text{ion}} \right\rangle $ as a single parameter by incorporating \( f_{\text{esc}} \) into \( \xi_{\text{ion}} \) \citep{Dayal:2018hft}. Following \citet{Price:2016uyw}, we apply a uniform prior on \( \log_{10} \xi_{\text{ion}} \in \mathcal{U}[23.5, 27.5] \) in units of \( \log_{10} \left[ \text{Hz} \, \text{erg}^{-1} \right] \). The clumping factor is initially treated as a free parameter with a uniform prior, setting an upper bound at $C_{\rm HII} \leq 10$. Later on, it is kept fixed at $C_{\rm HII}=5$, similar to \citet{Krishak:2021fxp}. This helps us explore the outcome of both cases for a comparative analysis.

In the final stage, we modify the public version of the Boltzmann solver code \texttt{CLASS} \citep{Blas_2011}, where this reconstructed reionization history is supplied as an input within the \texttt{thermodynamics.c} module, in place of the Planck $\tanh$ reionization model. This helps us consistently overcome any dependence of the baseline reionization model on the estimated parameters and search for possible consequences of the present learning method. For this we undertake a joint MCMC analysis on the 6 $\Lambda$CDM cosmological parameters and 2 reionization astrophysical parameters using \texttt{MontePython} \citep{Audren:2012wb,Brinckmann:2018cvx} by generating mock data for the upcoming 21-cm SKA mission along with some other data sets mentioned in Sec. \ref{sec:data}. We subsequently analyze the errors and correlations of the different model parameters. 

{
\subsection{Advancements over previous works}

Our work advances from the earlier works in this direction (see, for example, \citet[]{Krishak:2021fxp, Chatterjee:2021ygm, Paoletti:2021gzr, Adak:2024urf})  by multiple folds:

\begin{itemize}[left=0pt]
    \item We employ GPR by simultaneously training the parameters governing the GP mean function and the kernel hyperparameters (instead of fixing the mean function to the best-fit values) to obtain the predicted UV LF density profile.  
    \item Along with other astrophysical parameters, we keep the clumping factor \(C_{\rm HII}\) as a free parameter in MCMC analysis (and compare with earlier studies with a fixed value \(C_{\rm HII}\) = 5). 
    \item We consider both HFF and JWST as the reconstruction training data sets (along with possible combinations of other data sets), followed by a thorough, methodical comparative analysis between their role in constraining the astrophysical parameters and hence in deriving reionization history. 
    \item We extend our analysis to the yet-unexplored directions on the applications of GPR in reionization. This is materialized by considering, separately, the global 21-cm signal (from SARAS) and 21-cm power spectrum (from SKA), which helps explore their role in inferring the reionization physics. 
    \item And finally, we modify the Boltzmann solver code \texttt{CLASS} \citep{Blas_2011} to accommodate our reconstructed reionization history into the MCMC code \texttt{MontePython} \citep{Audren:2012wb,Brinckmann:2018cvx}. 
\end{itemize}
Thus, our findings are expected to have important implications for understanding the nature and distribution of the first light sources and their role in shaping the early Universe. 
}

\section{Results \& Discussions \label{sec:results}}

\begin{figure}
\centering 
    \begin{minipage}{\linewidth}
        \centering
        \includegraphics[width=0.6\linewidth]{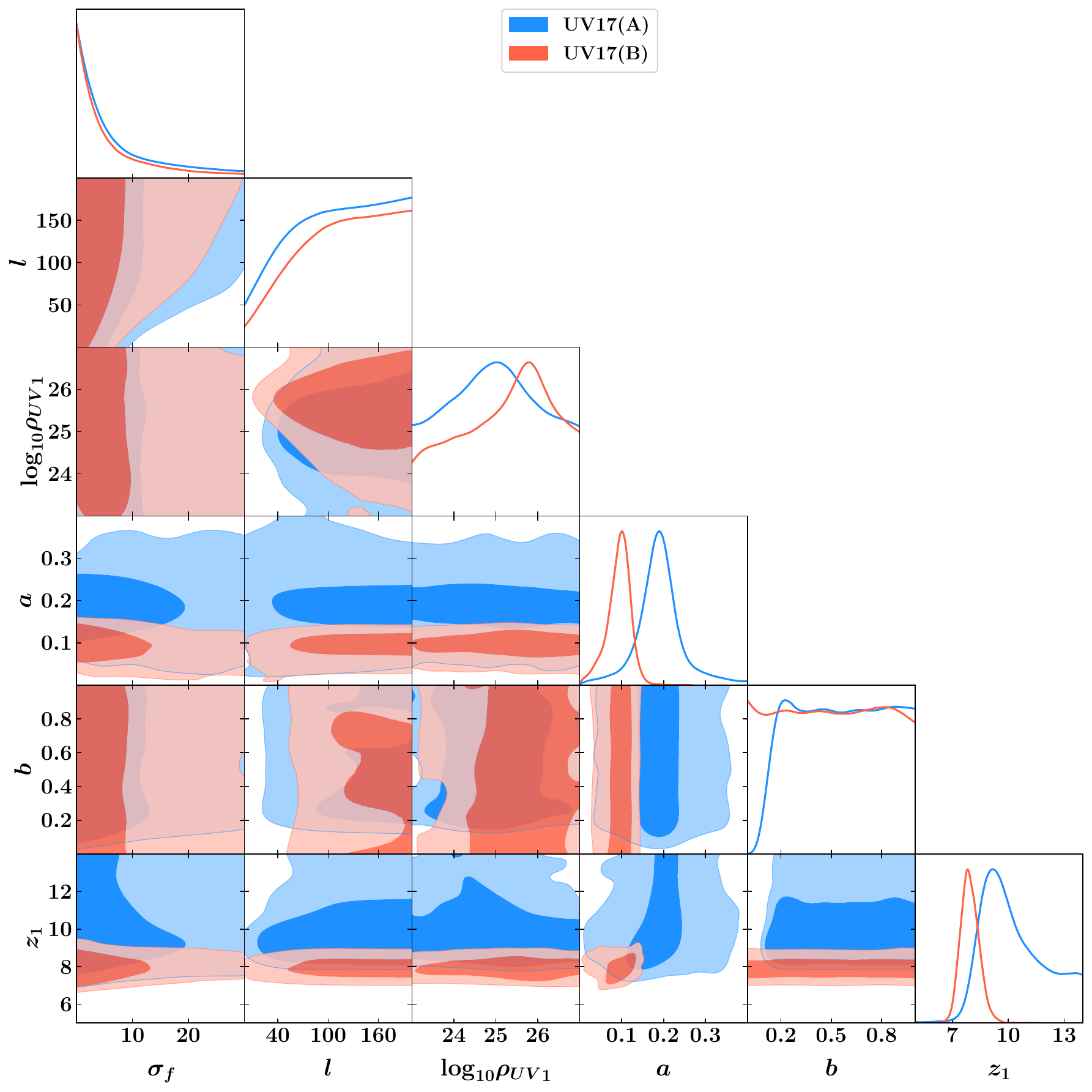} \\
        {(a) Trained MCMC samples of kernel hyperparameter and parameters governing the mean function for GPR \\[1.5cm] }
        \end{minipage}
 \begin{minipage}{\linewidth}
        \centering
\begin{minipage}{\linewidth}
\centering
\begin{minipage}{0.4\linewidth}
\centering 
\includegraphics[width=\linewidth]{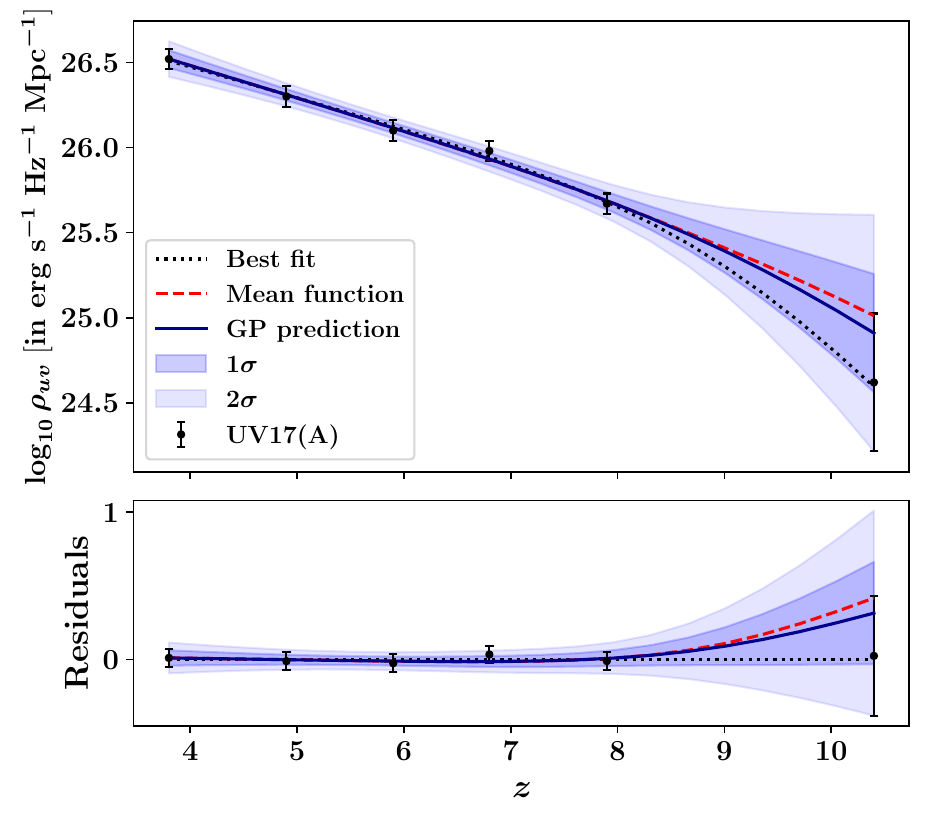}  \\
{(b) Reconstruction with UV17(A) data} 
\end{minipage}
\begin{minipage}{0.4\linewidth}
\centering
\includegraphics[width=\linewidth]{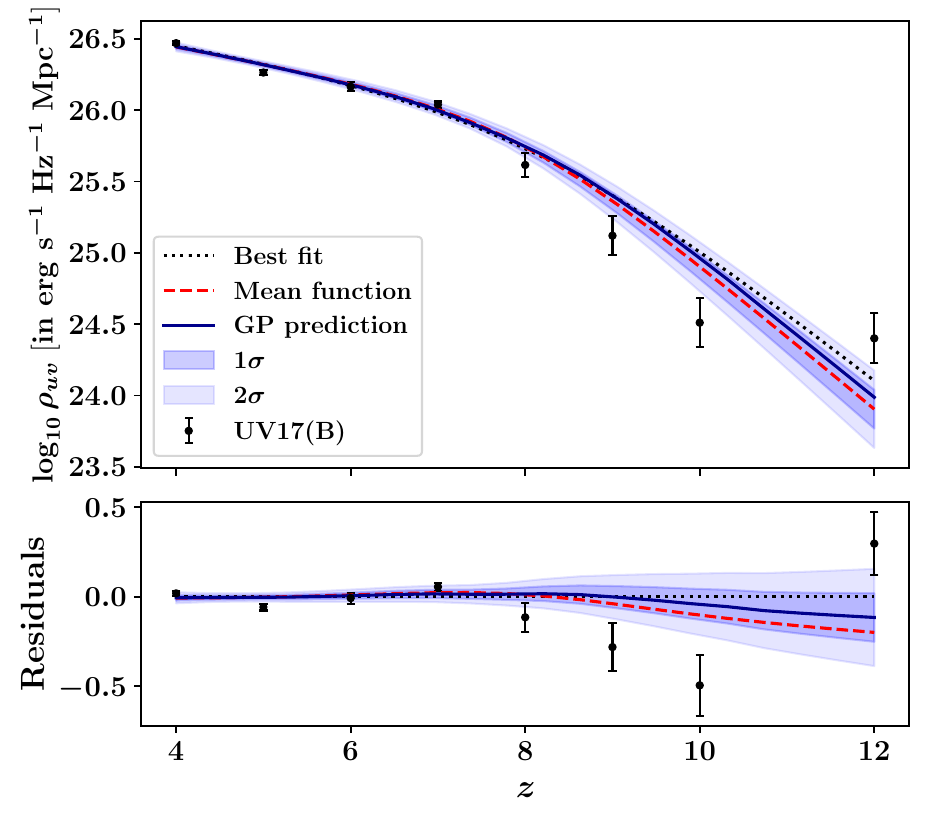}  \\
{(c) Reconstruction with UV17(B) data} 
\end{minipage}        
\end{minipage}
\end{minipage}
\caption{The reconstructed UV luminosity density function in the redshift range $z \sim 4-12$ obtained from Gaussian Process regression using the UV17(A) and UV17(B) data.}
\label{fig:rho_uv}
\end{figure}

\subsection{Analysis with existing datasets}

Following the methodology described in Sec. \ref{sec:uvlum}, we reconstruct the UV luminosity density as a function of redshift in the range $4<z<12$ employing GPR on the UV17(A) and UV17(B) data sets. Fig. \ref{fig:rho_uv}(b) and (c) illustrate the reconstructed UV luminosity density profile as a function of redshift for both UV17(A) and UV17(B) compilation, respectively. The solid blue curve represents the GP reconstructed mean curve, and the shaded regions correspond to the 1$\sigma$ and 2$\sigma$ uncertainties associated with the reconstructed curve. The black dotted lines give the best-fit curves assuming the logarithmic double power-law parametric form to model the $\log_{10} \rho_{\rm UV}$ data. The predicted logarithmic double power-law evolution as a mean function for GPR is shown with red dashed lines. Our findings indicate that the logarithmic double power law model is consistent with the reconstructed GP function and the UV17(A) and UV17(B) data within the redshift range $4<z<8$.  For $z>9$, the mean reconstructed curve deviates from the best-fit values. However, this deviation is included within 1$\sigma$ for the UV17(A) data and 2$\sigma$ for the UV17(B) data. In the case of UV17(B) data, the reconstructed UV luminosity curve excludes the $z=9$ and $z=10$ JWST $\log_{10} \rho_{\rm UV}$ data points. 

\begin{figure}
\centering
\begin{minipage}{0.49\linewidth}
\centering
\includegraphics[width=\textwidth]{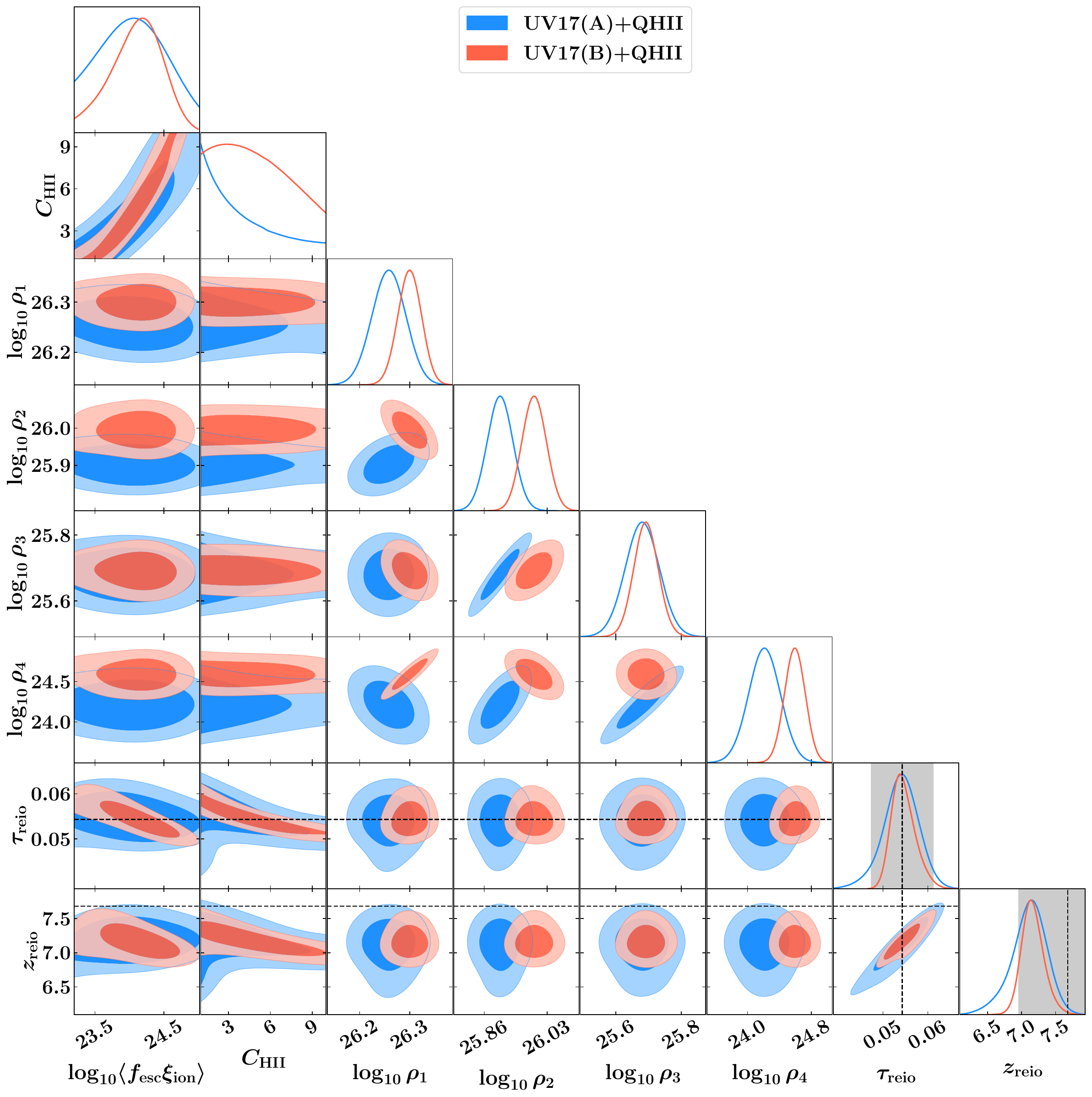} \\
(a)
\end{minipage}
\hfill
\begin{minipage}{0.49\linewidth}
\centering
\includegraphics[width=\textwidth]{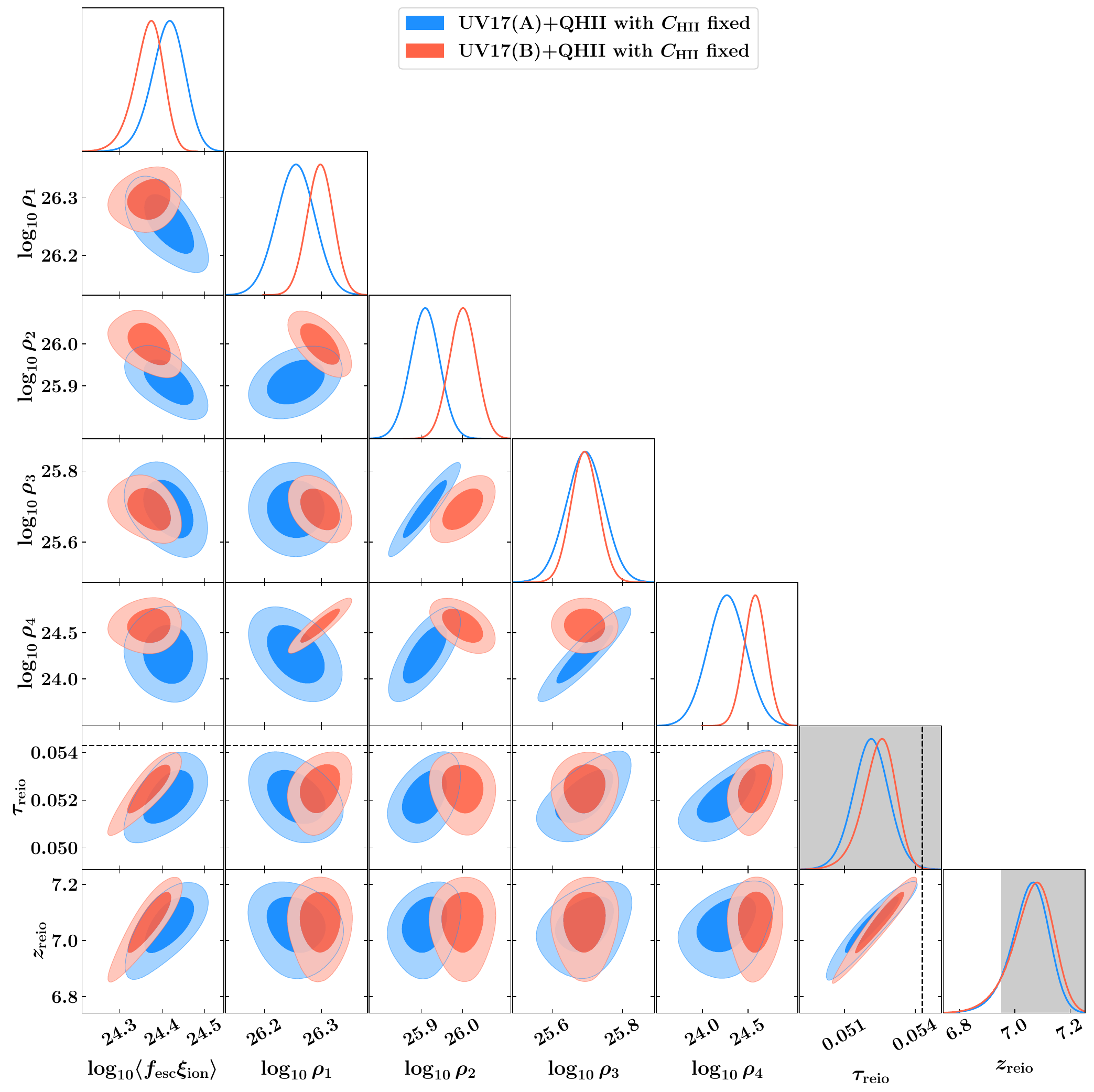} \\
(b)
\end{minipage} \\
\begin{minipage}{0.49\linewidth}
\centering
\includegraphics[width=\textwidth]{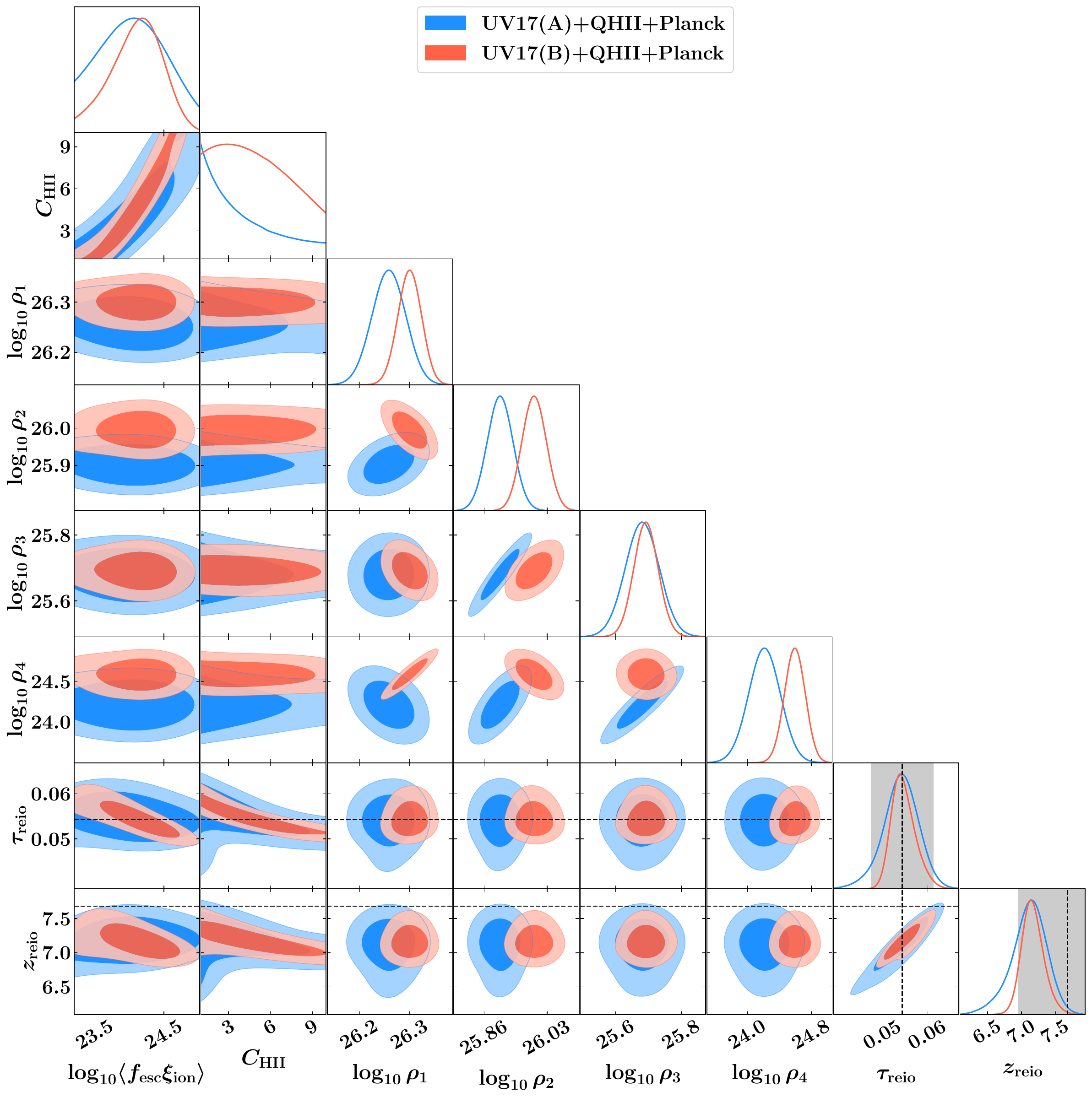} \\
(c)
\end{minipage}
\hfill
\begin{minipage}{0.49\linewidth}
\centering
\includegraphics[width=\textwidth]{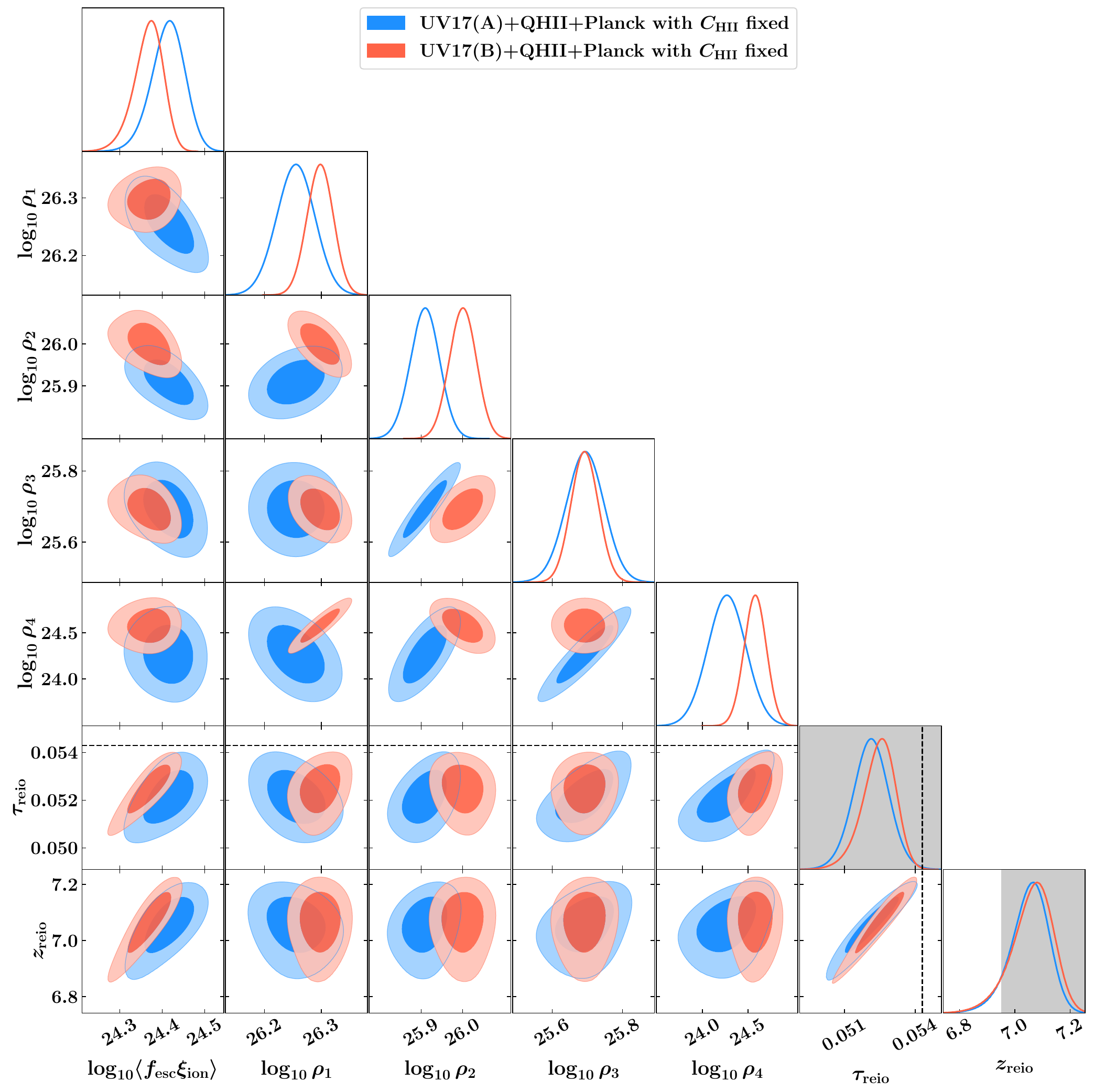} \\
(d)
\end{minipage}
\caption{Comparison between the constraints obtained on the astrophysical parameters employing the (a) UV17(A)+QHII vs UV17(B)+QHII, (b) with the clumping factor kept constant at $C_{\rm HII}$=5, (c) UV17(A)+QHII+Planck vs UV17(B)+QHII+Planck, and (d) with the clumping factor kept constant at $C_{\rm HII}$=5, during the MCMC. The black dashed line and shaded region represent the Planck best-fit with 1$\sigma$ CL.}
\label{fig:mcmc_present}
\end{figure}

\begin{figure}
    \centering
    \begin{minipage}{0.49\linewidth}
        \centering
        \includegraphics[width=\linewidth]{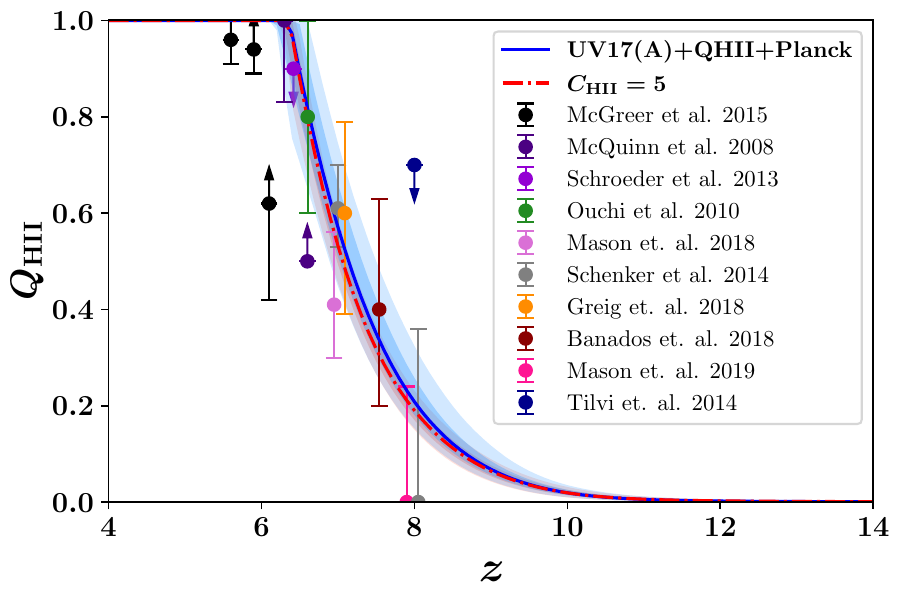}\\
        {(a)}
    \end{minipage}
    \hfill
    \begin{minipage}{0.49\linewidth}
        \centering
        \includegraphics[width=\linewidth]{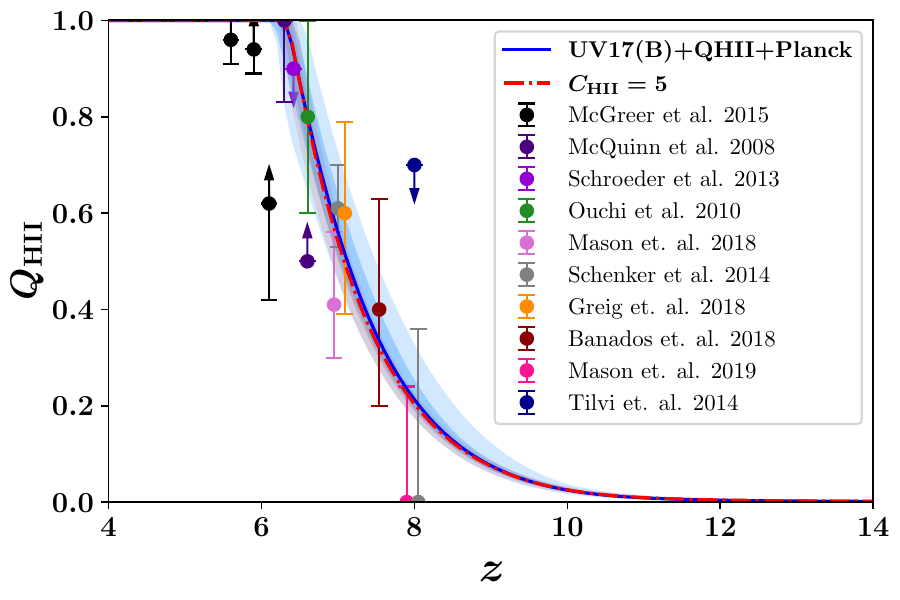}\\
        {(b)}
    \end{minipage} \\
    \begin{minipage}{0.49\linewidth}
        \centering
        \includegraphics[width=\linewidth]{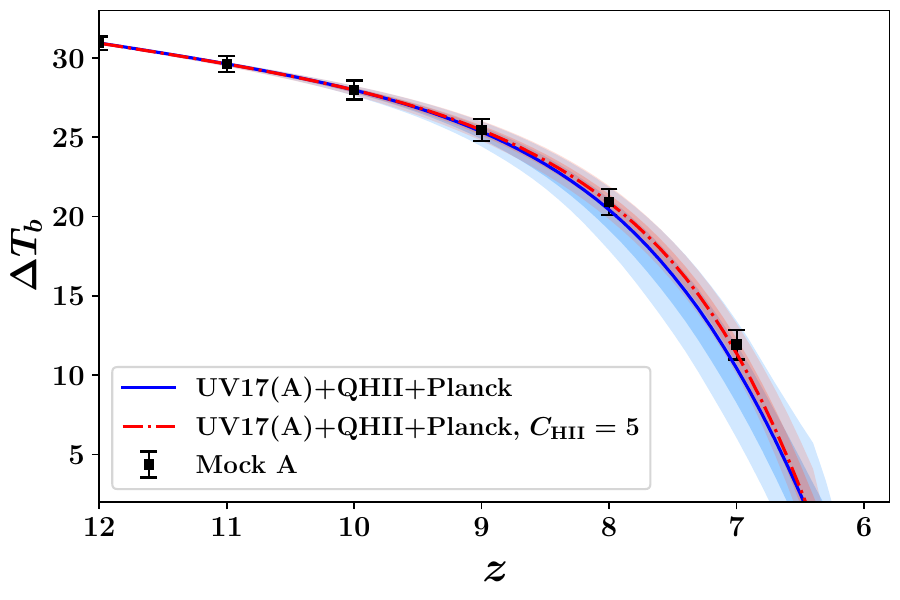}\\
        {(c)}
    \end{minipage}
    \hfill
    \begin{minipage}{0.49\linewidth}
        \centering
        \includegraphics[width=\linewidth]{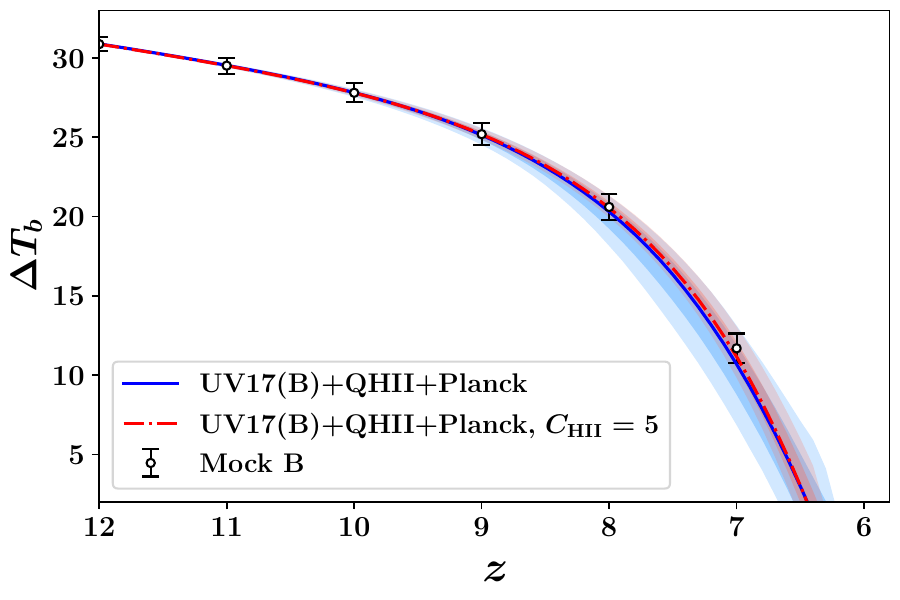}\\
        {(d)}
    \end{minipage}
    \caption{Evolution of the ionization fraction ($Q_{\rm HII}$) and global 21-cm signal ($\Delta T_b$) from reconstruction. Plots (a) and (b) show ionization fraction obtained using UV17+QHII+Planck (blue curve) vs UV17+QHII+Planck for a fixed value of the clumping factor  $C_{\rm HII}=5$ (red dashed line) from the GP reconstruction of the UV luminosity density with the UV17(A) (left panels) and UV17(B) (right panels) compilation. Plots (c) and (d) show the global 21-cm signal at the reionization epoch from respective data sets. {The QHII data is plotted with references for comparison. Mock A and B represent the simulated $\Delta T_b$ vs $z$ catalogues.}}
    \label{fig:qhii_Tb_plots}
\end{figure}

With this reconstructed UV luminosity profile, we proceed to trace the evolution of the ionization fraction $Q_{\rm HII}$, following the methodology described in Sec. \ref{sec:reion}. For this we adopt four equidistant redshift nodes at $z_1=4$, $z_2=6$, $z_3=8$ and $z_4=10$, where the values of logarithmic UV luminosity densities are redefined as $\log_{10} \rho_1$, $\log_{10} \rho_2$, $\log_{10} \rho_3$ and $\log_{10} \rho_4$ respectively. We make use of Eq. \eqref{eq:ioneqn} to solve for $Q_{\rm HII}$ where the reconstructed values of the UV luminosity densities (at the 4 redshift nodes) are treated as free parameters during the Bayesian MCMC analysis. The entire exercise is undertaken employing different data sets, mentioned in Sec. \ref{sec:data}, namely, UV17(A) \& UV17(B) in combination with QHII Ly$\alpha$ and Planck $\tau_{\rm reio}$ data. Fig. \ref{fig:mcmc_present}(a)-(b) shows the 2D-confidence contours and 1D-marginalized posteriors for the relevant parameters upon MCMC done using the joint UV17(A)+QHII (in blue) and UV17(B)+QHII (in red) data sets. {The black dashed line and shaded region represent the Planck baseline best-fit with 1$\sigma$ CL.} Similarly, Fig. \ref{fig:mcmc_present}(c)-(d) depicts the same for UV17(A)+QHII+Planck (in blue) and UV17(B)+QHII+Planck (in red) combinations.  It should be noted that, in panels (a) and (c) of Fig. \ref{fig:mcmc_present}, the clumping factor $C_{\rm HII}$ is treated as a free parameter during MCMC, whereas the panels (b) and (d) shows the case when the value of the clumping factor is kept fixed at $C_{\rm HII} = 5$ respectively. For both (a) and (c), we find that the parameters $\log_{10} \left\langle f_{\rm esc} \xi_{\rm ion} \right\rangle$ and $C_{\rm HII}$ are positively correlated. This feature indicates an anti-correlation with $t_{\rm rec}$, i.e., a reduction in $\left\langle f_{\rm esc} \xi_{\rm ion} \right\rangle$ lowers the source term, which is only offset by a longer recombination period to maintain ionization [see \citet[]{Gorce:2017glg, 2019MNRAS.485.3947M, Paoletti:2021gzr, Krishak:2021fxp}]. The values of UV luminosity densities at redshift points $z_1$, $z_2$, and $z_4$ differ greatly between the two different combinations of the data sets. However, at $z_3$, the average values are very similar, where the inclusion of early JWST data in the UV17(B) compilation narrows down the range of possible values. We also find that the nature of correlations between the parameters $\log_{10} \rho_{1-4}$ remains unchanged on fixing the value of $C_{\rm HII}$ (for comparison see panels (a) and (c) vs (b) and (d) of Fig. \ref{fig:mcmc_present}).

\begin{figure}
    \centering
    \begin{minipage}{0.49\linewidth}
        \centering
        \includegraphics[width=\linewidth]{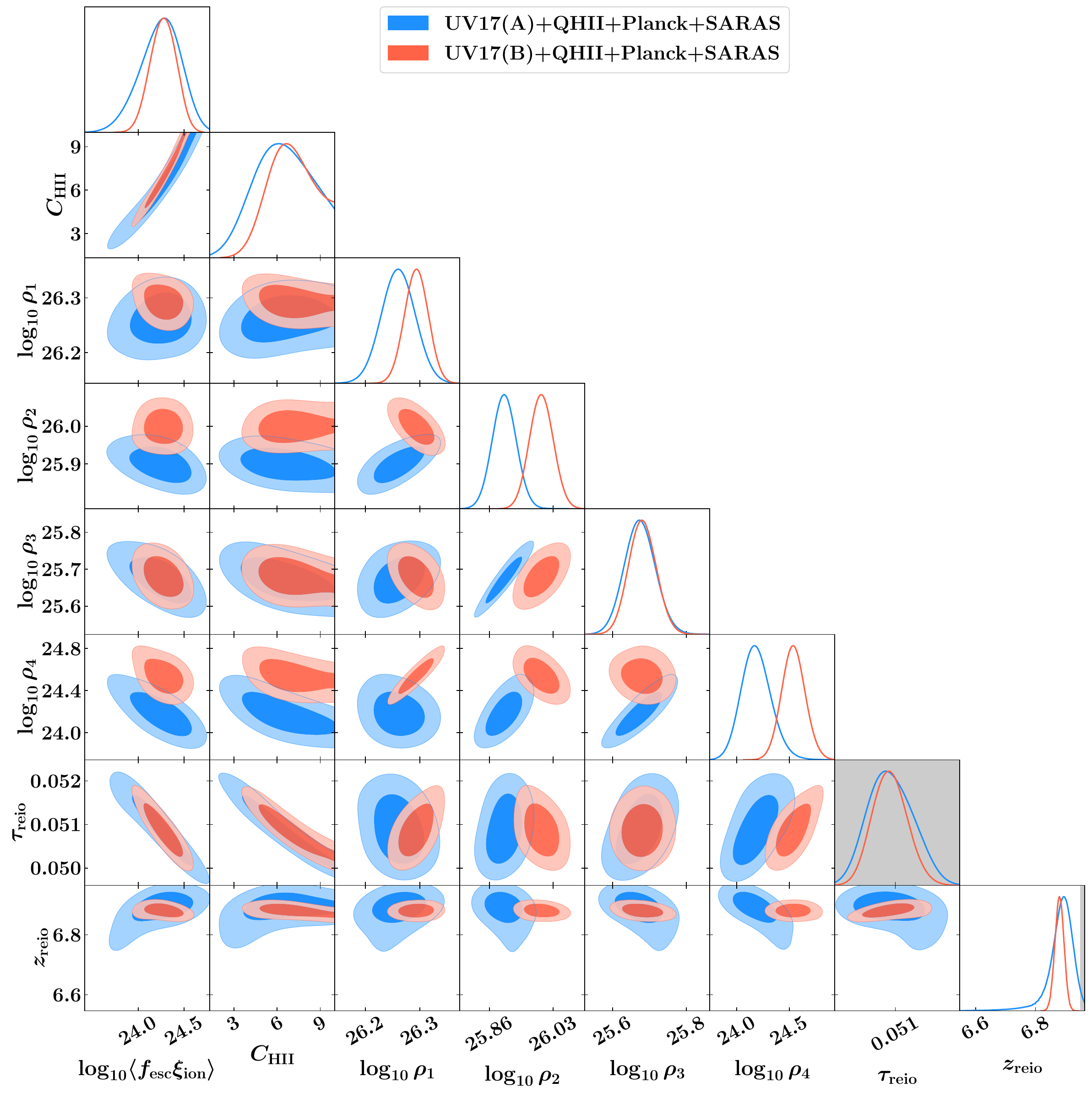}\\
        {(a)}
    \end{minipage}
    \hfill
    \begin{minipage}{0.49\linewidth}
        \centering
        \includegraphics[width=\linewidth]{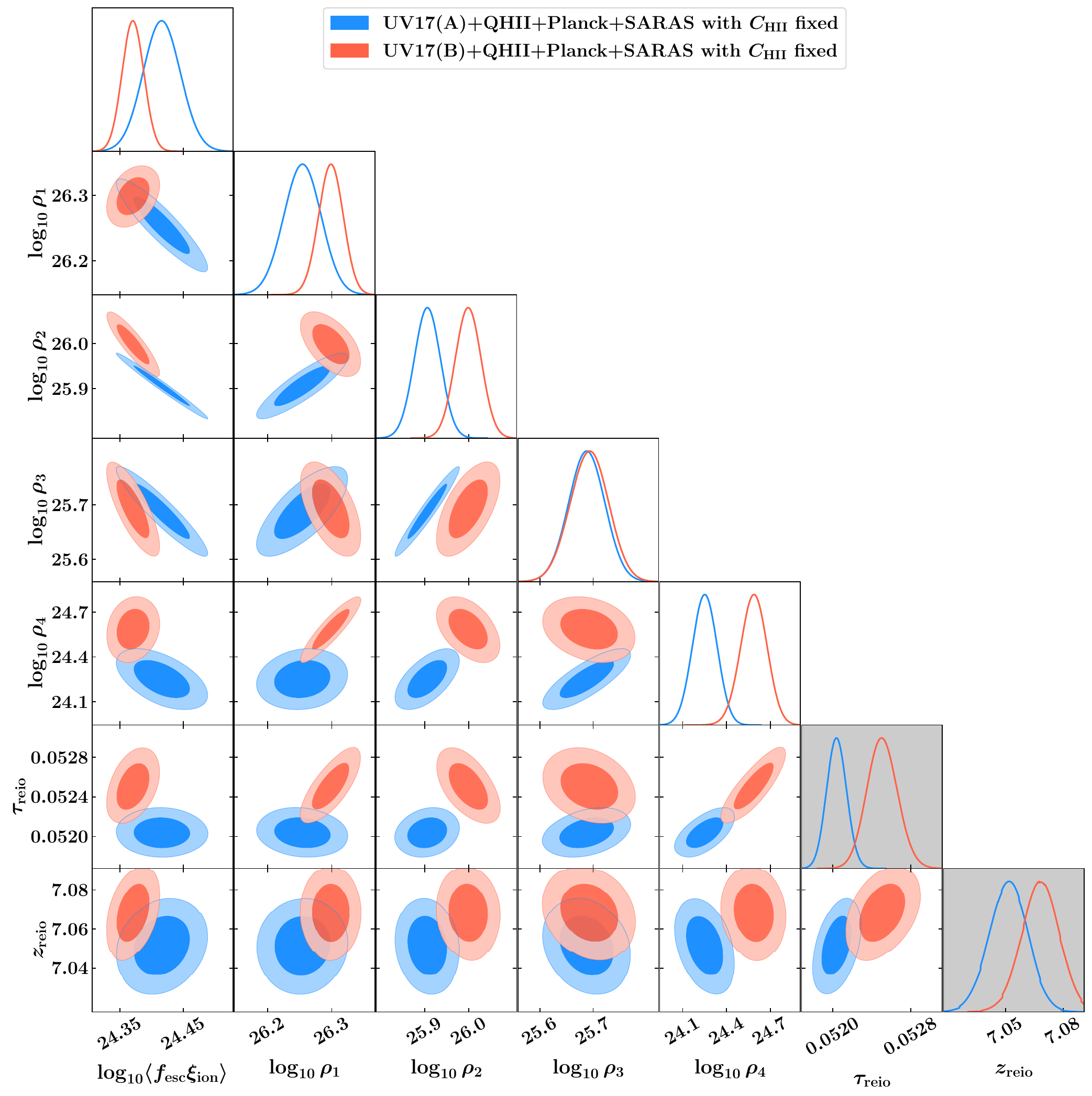}\\
        {(b)}
    \end{minipage} \\
    \begin{minipage}{0.49\linewidth}
        \centering
        \includegraphics[width=\linewidth]{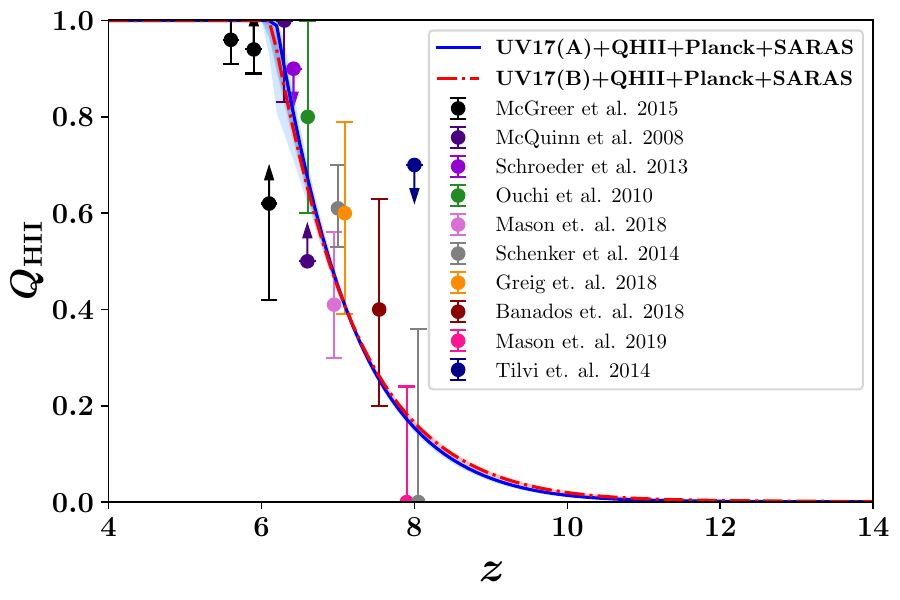}
        {(c)}
    \end{minipage}
    \hfill
    \begin{minipage}{0.495\linewidth}
        \centering
        \includegraphics[width=\linewidth]{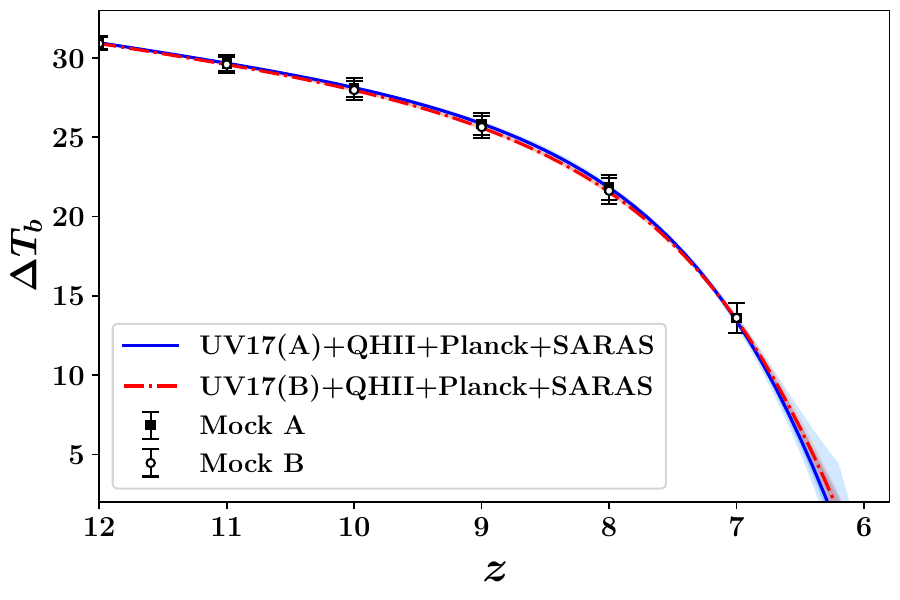}
        {(d)}
    \end{minipage} 
    \caption{Comparison between the bounds obtained on the astrophysical parameters using UV17(A)+QHII+Planck+SARAS vs UV17(B)+QHII+Planck+SARAS and (b) with a constant clumping factor of $C_{\rm HII}=5$ (panel (b)). Evolution of the ionization fraction in (c) and the global 21-cm signal at reionization epoch in (d) obtained using UV17(A)+QHII+Planck+SARAS (blue curve) and UV17(B)+QHII+Planck+SARAS (red dashed line) from the GP reconstruction of the UV luminosity density with the UV17(A) and UV17(B) compilations respectively}
    \label{fig:saras}
\end{figure}

We are now in a position to reconstruct the reionization history profile with the help of these obtained bounds on the parameters $\log_{10} \left\langle f_{\rm esc} \xi_{\rm ion} \right\rangle$, $C_{\rm HII}$, and $\log_{10} \rho_{1-4}$, and by deriving the evolution of the ionization fraction via Eq. \eqref{eq:ioneqn}. Fig. \ref{fig:qhii_Tb_plots} demonstrates the comparison between the reconstructed $Q_{\rm HII}$ using two different data compilations - UV17(A)+QHII+Planck in panel (a) vs UV17(B)+QHII+Planck in panel (b). With this reconstructed $Q_{\rm HII}$ profile, we can trace the nature of the 21-cm global brightness temperature fluctuation $\Delta T_b$, directly employing Eq. \eqref{eq:deltaTb}. Plots for the reconstructed $\Delta T_b$ for the UV17(A)+QHII+Planck and UV17(B)+QHII+Planck data sets are shown in the panels (c) and (d) of Fig. \ref{fig:qhii_Tb_plots} respectively. As depicted in Fig. \ref{fig:qhii_Tb_plots}, we obtain more or less similar results for the reionization history profile and global signal employing Planck+$Q_{\rm HII}$+ either UV17(A) or UV17(B) data sets. Fixing the clumping factor $C_{\rm HII}$ to a constant value (here $C_{\rm HII}=5$) results in a more precise reconstruction of both $Q_{\rm HII}$ and $\Delta T_b$. This feature is apparent from Fig. \ref{fig:mcmc_present}(b) and \ref{fig:mcmc_present}(d), where we notice a significant reduction of the $\left\langle f_{\rm esc} \xi_{\rm ion} \right\rangle$ parameter space obtained with MCMC analysis. However, one should take it with a pinch of salt, as there is no a priori reason as to why $C_{\rm HII}$ has to take a fixed value, especially when the parameters are not indifferent to a running $C_{\rm HII}$ and hence, the constraints obtained by keeping $C_{\rm HII}$ free should be more acceptable in a conservative approach.

\subsection{Forecasting on upcoming surveys}

Let us now engage ourselves in investigating the role of the global 21-cm signal on this reconstruction. After plotting the evolution of the 21-cm global signal with best-fit astrophysical parameters from UV17(A)+QHII+Planck and UV17(B)+QHII+Planck data, we create mock global 21-cm signal $\Delta T_b$ vs $z$ data {(shown in Fig. \ref{fig:qhii_Tb_plots}(c)-(d))}, assuming the instrumental specifications of SARAS. We then redo this entire exercise of learning the reionization history incorporating the SARAS mock data in combination with the remaining UV17(A)/UV17(B), QHII and Planck data sets. Similar to the previous plots, Fig. \ref{fig:saras}(a)-(b) shows the 2D-confidence contours and 1D-marginalized posteriors for the MCMC parameter space using the joint UV17(A)+QHII+Planck+SARAS and UV17(B)+QHII+Planck+SARAS data sets, respectively. In panel (a) of Fig. \ref{fig:saras}, the clumping factor $C_{\rm HII}$ is treated as a free parameter during MCMC, whereas, panel (b) shows the case when the value of the clumping factor is kept fixed at $C_{\rm HII} = 5$ respectively.  Therefore, the inclusion of SARAS data, as shown in Fig. \ref{fig:saras}, provides much tighter bounds on the astrophysical parameters. We also plot the evolution of the ionization fraction and global 21-cm signal at the reionization epoch, in the Fig. \ref{fig:saras}(c) and Fig. \ref{fig:saras}(d), respectively. Our findings show that the reconstructed $Q_{\rm HII}$ and $\Delta T_b$ profiles are now further constrained, compared to the previous cases. Thus, one can conclude that although the UV17(A)+QHII+Planck and UV17(B)+QHII+Planck data influence the reionization history and bounds on the astrophysical parameters, the global differential brightness temperature has a more significant impact on these parameters, thereby helping to learn the cosmic reionization history with better precision.

\begin{table}
{\renewcommand{\arraystretch}{1.15} \setlength{\tabcolsep}{10 pt} \centering
\begin{tabular}{l   c   c    c   c   c}
\hline \hline
\textbf{Data sets} & $\boldsymbol{\log_{10} \langle f_{\rm esc} \, \xi_{\rm ion} \rangle }$ &  $\boldsymbol{C_{\rm HII}}$ &  $\boldsymbol{\tau_{\rm reio}}$ & $\boldsymbol{z_{\rm reio}}$ & $\boldsymbol{\Delta z}$   \\
    \hline
    UV17(A)+QHII     &  $24.001_{-0.853}^{+0.405}$  &   $3.302_{-1.967}^{+3.648}$  &  $0.054_{-0.0026}^{+0.0038}$  & $7.12_{-0.19}^{+0.21}$  &   $2.21_{-0.19}^{+1.23}$  \\ 
    UV17(A)+QHII, $C_{\rm HII} = 5$     &   $24.417^{+0.032}_{-0.039}$   &   5    &  $0.052_{-0.0007}^{+0.0007}$   &   $7.06_{-0.07}^{+0.05}$   &  $2.15_{-0.12}^{+0.14}$   \\ 
    UV17(A)+QHII+Planck     &   $24.192_{-0.411}^{+0.281}$   &   $4.582_{-2.191}^{+3.197}$    &  $0.054_{-0.0019}^{+0.0033}$   &   $7.13_{-0.11}^{+0.19}$  &  $2.17_{-0.15}^{+0.24}$  \\ 
    UV17(A)+QHII+Planck, $C_{\rm HII} = 5$  &   $24.416_{-0.040}^{+0.034}$   & 5 &  $0.052_{-0.0007}^{+0.0007}$ &  $7.06_{-0.07}^{+0.05}$    &   $2.15_{-0.13}^{+0.14}$   \\ 
    UV17(A)+QHII+Planck+SARAS   &  $24.192_{-0.411}^{+0.281}$  &  $ 4.582_{-2.191}^{+3.197}$  &  $0.054_{-0.0019}^{+0.0033}$  & $7.13_{-0.11}^{+0.19}$ &  $2.17_{-0.15}^{+0.24}$  \\ 
    UV17(A)+QHII+Planck+SARAS, $C_{\rm HII} = 5$  &   $24.416_{-0.028}^{+0.027}$   &  5  &   $0.052_{-0.0001}^{+0.0001}$  &  $7.05_{-0.01}^{+0.01}$ &  $2.13_{-0.05}^{+0.06}$   \\ 
    \hline
    UV17(B)+QHII   &  $24.164_{-0.458}^{+0.262}$  &  $4.760_{-2.485}^{+2.978}$  &  $0.054_{-0.0015}^{+0.0034}$ &  $7.13_{-0.10}^{+0.18}$  &  $2.25_{-0.07}^{+0.20}$ \\ 
    UV17(B)+QHII, $C_{\rm HII} = 5$  &  $24.373_{-0.037}^{+0.026}$  &  5  &  $0.053_{-0.0007}^{+0.0005}$  &  $7.08_{-0.09}^{+0.04}$ &  $2.24_{-0.04}^{+0.05}$  \\ 
    UV17(B)+QHII+Planck     &   $24.200_{-0.441}^{+0.247}$   &  $5.081_{-2.610}^{+3.027}$  &  $0.054_{-0.0013}^{+0.0031}$  &  $7.12_{-0.10}^{+0.17}$  & $2.25_{-0.08}^{+0.15}$ \\ 
    UV17(B)+QHII+Planck, $C_{\rm HII} = 5$  &  $24.374_{-0.036}^{+0.025}$   &  5  &  $0.053_{-0.0007}^{+0.0005}$   & $7.08_{-0.08}^{+0.05}$  & $2.25_{-0.05}^{+0.06}$  \\ 
    UV17(B)+QHII+Planck+SARAS     &  $24.273_{-0.141}^{+0.145}$   &  $6.723_{-1.304}^{+1.933}$  &  $0.051_{-0.0004}^{+0.0004}$   & $6.88_{-0.01}^{+0.01}$  & $2.24_{-0.03}^{+0.04}$  \\ 
    UV17(B)+QHII+Planck+SARAS, $C_{\rm HII} = 5$  &  $24.370_{-0.016}^{+0.017}$  &   5   &  $0.053_{-0.0001}^{+0.0001}$  &  $7.07_{-0.01}^{+0.01}$  &   $2.25_{-0.04}^{+0.03}$ \\ 
    \hline
    \hline
\end{tabular}
\caption{Summary of the mean and $1\sigma$ bounds  obtained on the astrophysical parameters and reionization history using different data compilations.}
\label{tab:summary}
}
\end{table}

The constraints on the astrophysical parameters  $\log_{10} \langle f_{\rm esc} \xi_{\rm ion} \rangle, C_{\rm HII}$, optical depth \( \tau_{\rm reio} \), reionization redshift $z_{\text{re}}$ and reionization duration $\Delta z$, separately for each individual data compilations explored in the present analysis, are summarized in Table \ref{tab:summary}. The table shows that the optical depth constraints from all the combinations - UV17+QHII, UV17+QHII+Planck and UV17+QHII+Planck+SARAS align with the 1$\sigma$ optical depth values from Planck 2018 results. Our analysis of reionization duration, \( \Delta z \), suggests that a substantial portion of reionization (from 10\% to 90\% ionization) occurs over approximately 2 units (for Planck+UV17+QHII). The 68\% and 95\% confidence intervals reveal that the marginalized posterior distribution of $\Delta z$ is slightly skewed when the clumping factor is free to vary during the MCMC. The redshift at which reionization reaches 50\% completion, denoted as \( z_{\text{reio}} \), is found to be approximately 7 for all the above-mentioned data combinations. The table shows that the mean value of $\log_{10} \langle f_{\rm esc} \xi_{\rm ion} \rangle$ parameter is approximately 24 units from all combinations of data sets. With the inclusion of SARAS data and keeping $C_{\rm HII}$ fixed, the 1$\sigma$ bounds on the $\log_{10} \langle f_{\rm esc} \xi_{\rm ion} \rangle$ parameter are significantly constrained. The $C_{\rm HII}$ parameter takes different values for different data set combinations. The 1$\sigma$ bound on this parameter improves with the UV17(B)+QHII+Planck+SARAS data set combination.

\begin{figure}
    \centering
    \includegraphics[width=\textwidth]{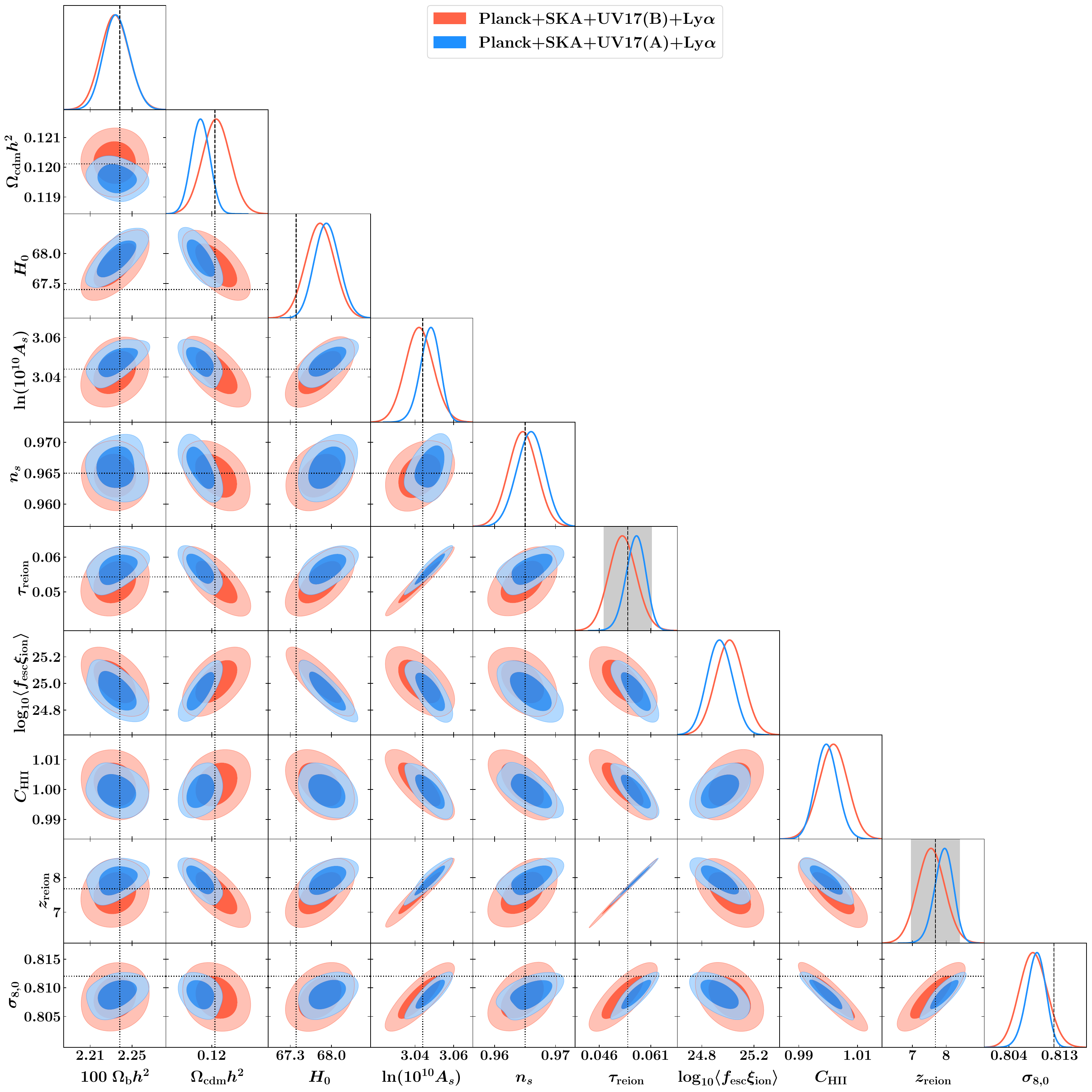}
   \caption{Forecasting the reionization power spectrum of 6 $\Lambda$CDM parameters with 2 astrophysical parameters: ionization efficiency and clumping factor, using the Planck+SKA+Ly$\alpha$+UV17(A) data sets and the Planck+SKA+Ly$\alpha$+UV17(B) data sets. Here for this forecast and MCMC analysis, we have used fake Planck realistic data. The analysis is undertaken using the mean reconstructed $Q_{\rm HII}$ function, without incorporating the errors. }
    \label{fig:MCMCPk}
\end{figure}

Our final target is to explore the prospects of the upcoming 21-cm mission SKA in simultaneously inferring the 2 astrophysical + 6 cosmological parameters. For this, we make use of the reconstructed reionization history and compute the 21-cm power spectrum during the reionization epoch with a conservative approach, employing Eq. \eqref{eq:P21}. Further, we modify the Boltzmann solver code \texttt{CLASS} in order to accommodate our reconstructed reionization history (as opposed to using baseline Planck $\tanh$ reionization model), as explained in Sec. \ref{sec:reion}. The underlying cosmological model was set to the standard 6-parameter $\Lambda$CDM framework. Following the prescription given by \citet[]{Sprenger:2018tdb}, we generate mock catalogues for the future SKA mission in the reionization era between redshift $z \approx 6-12$, from the simulated Planck realistic data, utilizing the \texttt{fake\_planck\_realistic} likelihood in \texttt{MontePython} (hereafter referred to as Planck) and MIKE-HIRES Lyman-$\alpha$ forest dataset, adopting the fiducial values of the cosmological parameters as $w_{\rm b}=0.02237$, $w_{\rm cdm}=0.12010$, $\ln[10^{10}A_s]=3.0447$, $n_s=0.9659$, $H_0=67.8$ km Mpc$^{-1}$ s$^{-1}$, $\tau_{\rm reio}=0.0543$, consistent with Planck 2018 data \citep[]{refId0}. Finally, we undertake a Bayesian MCMC analysis to forecast the 6 cosmological and 2 astrophysical parameters using the \texttt{MontePython} code. We adopt uniform priors for all parameters as: $100~ \Omega_{\rm b}h^2 \in \mathcal{U}[0.05,10]$, $\Omega_{\rm c}h^2 \in \mathcal{U}[0.01,0.99]$, $H_0 \in \mathcal{U}[50,100]$, $n_s \in \mathcal{U}[0.5,1.5]$, $\tau_{\rm reio} \in \mathcal{U}[0.004,0.2]$, $\ln \left( 10^{10} A_s \right) \in \mathcal{U}[1, 5]$, $\left\langle f_{\rm esc} \xi_{\rm ion} \right\rangle \in \mathcal{U}[20,30]$, and $C_{\rm HII} \leq 10$ respectively. The resulting  2D-confidence contours and 1D-marginalized posteriors for the Planck+SKA+Ly$\alpha$+UV17(A) and Planck+SKA+Ly$\alpha$+UV17(B) data sets have been presented in Fig. \ref{fig:MCMCPk}, which also helps us in easy comparison between the two data sets.  Table \ref{tab:summary-Pk} presents the mean and $1\sigma$ bound on the astrophysical and cosmological parameters obtained from this analysis. Besides, we also show the results for two derived parameters $z_{\rm reio}$ and $\sigma_{8,0}$.

\begin{table}
{\renewcommand{\arraystretch}{1.15} \setlength{\tabcolsep}{9 pt} \centering
\begin{tabular}{l  |  c   | c  |  c }
\hline \hline
\multirow{ 2}{*}{\bf Parameters} & $\tanh$ reionization model \footnote{The values of the astrophysical parameters are taken from \citep[]{Barkana:2000fd, Mesinger:2016ddl, Price:2016uyw, Sarkar:2015dib, 2018JCAP...08..045D, Dayal:2018hft, Dey:2022ini}}  & \multicolumn{2}{c}{GP reconstructed reionization model} \\
\cline{2-4}
 & Planck 2018 TT,TE,EE + lowE & Planck+SKA+Ly$\alpha$+UV17(A) & Planck+SKA+Ly$\alpha$+UV17(B) \\
\hline
$\boldsymbol{\log_{10} \langle f_{\rm esc} \, \xi_{\rm ion} \rangle }$ & $23-500$ & $24.94_{-0.01}^{+0.0095}$ & $24.95_{-0.11}^{+0.093}$ \\
$\boldsymbol{C_{\rm HII}}$ & $1-10$ & $1.006_{-0.006}^{+0.0059}$ & $1.003_{-0.003}^{+0.0037}$ \\
$\boldsymbol{100 \,\Omega_{\rm b}h^2}$ & $2.236 \pm 0.015$ & $2.235_{-0.013}^{+0.0095}$ & $2.232_{-0.013}^{+0.01}$ \\
$\boldsymbol{\Omega_{\rm cdm} h^2}$ & $0.1202 \pm 0.0014 $ & $0.1196_{-0.00031}^{+0.00033}$ & $0.1197_{-0.0004}^{+0.0003}$ \\
$\boldsymbol{H_{0}}$ & $67.27 \pm 0.60  $ & $67.92_{-0.21}^{+0.19}$ & $67.9_{-0.22}^{+0.21}$ \\
$\boldsymbol{\ln \left(10^{10} A_{s} \right)}$ & $3.045 \pm 0.016 $ & $3.047_{-0.0045}^{+0.0066}$ & $3.045_{-0.0064}^{+0.0056}$ \\
$\boldsymbol{n_{s}}$ & $0.9649 \pm 0.0044 $ & $0.9658_{-0.0019}^{+0.0023}$ & $0.9656_{-0.002}^{+0.0024}$ \\ 
$\boldsymbol{\tau_{\rm reio}}$ & $0.0544_{-0.0081}^{+0.0070}$ & $0.0566_{-0.0025}^{+0.003}$ & $0.05595_{-0.0027}^{+0.0037}$ \\
$\boldsymbol{100 \, \theta_{s}}$ & $1.04077 \pm 0.00047 $ &  $1.044_{-0.00024}^{+0.00027}$ & $1.044_{-0.00029}^{+0.00027}$ \\
$\boldsymbol{z_{\rm reio}}$ & $7.68 \pm 0.79$ & $7.93_{-0.26}^{+0.28}$ & $7.865_{-0.26}^{+0.36}$ \\
$\boldsymbol{\sigma_{8}}$ & $0.8120 \pm 0.0073 $ & $0.8088_{-0.0013}^{+0.0019}$ & $0.8087_{-0.0015}^{+0.002}$ \\
\hline
\hline
\end{tabular}
\caption{Summary of the mean and $1\sigma$ bounds obtained on the astrophysical and cosmological parameters using the $\tanh$ model of reionization for Planck TT + lowE data \citep[]{refId0} and using the reconstructed reionization history for the Planck+SKA+Ly$\alpha$+UV17(A) and UV17(B) data sets. {Note that the allowed ranges of the two astrophysical parameters $\log_{10} \langle f_{\rm esc} \, \xi_{\rm ion} \rangle, C_{\rm HII}$ are derived from different astrophysical observations, without considering the $\tanh$ reionization model, while for the rest of the cosmological parameters, the allowed ranges are based on the $\tanh$ reionization model.}}
\label{tab:summary-Pk}
}
\end{table}

A comparison of the results presented in Table \ref{tab:summary} and Table \ref{tab:summary-Pk} for the two key cosmological parameters \(\tau_{\rm reio}\) and \(z_{\rm reio}\),  as well as the two astrophysical parameters \(C_{\rm HII}\) and \(\langle f_{\rm esc} \xi_{\rm ion} \rangle\) is in order. Based on the power spectrum analysis, we achieve tighter \(1\sigma\) bounds on the astrophysical parameters \(\log_{10} \langle f_{\rm esc} \, \xi_{\rm ion} \rangle\) and \(C_{\rm HII}\) compared to the results in Table \ref{tab:summary}. The parameter \(z_{\rm reio}\), which denotes the redshift at which 50\% ionization is complete, shows a relatively higher value from the power spectrum analysis (somewhat akin to $\tanh$ reionization model) than from the global signal reconstruction analysis. However, a crucial difference between the two analyses needs to be kept in mind. In the power spectrum analysis, we considered the full fake Planck realistic data and ran MCMC for the 6 cosmological parameters along with the astrophysical parameters. In contrast, for the global signal reconstruction, we used the Planck optical depth measurement and ran MCMC for the astrophysical parameters only. This difference may have shown up as the slight difference in the $z_{\rm reio}$ values estimated from the two analyses and one should rely more on the ``all parameters open'' case than the other one.

Further, for comparison with the baseline reionization model, we also present the results for the 6-parameter $\Lambda$CDM with $\tanh$ reionization model in Table \ref{tab:summary-Pk}, which shows that our obtained values of $\tau_{\rm reio}$, $z_{\rm reio}$, $H_{0}$, $\sigma_{8}$, and other cosmological parameters using the GP reconstrued reionization history are well consistent with the Planck 2018 + baseline model. On top of that, both of the astrophysical parameters \(C_{\rm HII}\) and \(\langle f_{\rm esc} \xi_{\rm ion} \rangle\) prefer values closer to the lower bounds of the tanh model. This happens in spite of choosing a wide prior range for both of them. We observe that the mean values of \( H_0 \) remain almost unaffected and lie close to the baseline Planck values, similar to \citet[]{Chatterjee:2021ygm}. However, the mean value of \(\sigma_8\) slightly decreases when changing the reionization history. This feature is similar to previous observations made in Fig. 4 of \citet[]{Hazra:2019wdn} using non-parametric nodal reconstruction, nonetheless that the astrophysical parameters were kept fixed to $\log_{10} \left\langle f_{\rm esc} \xi_{\rm ion} \right\rangle $ = 24.85 \citep[]{Ishigaki_2015} and $C_{\rm HII} \leq 3$, focusing on examining how the reconstructed model affects the constraints on the cosmological parameters. \citet[]{Paoletti:2021gzr} extended this exercise by varying the reionization astrophysical parameters simultaneously with the cosmological parameters using Planck+UV17+QHII data sets. The constraints obtained on the key parameters, such as $\log_{10} \left \langle f_{\rm esc} \xi_{\rm ion} \right\rangle$ and $\tau_{\rm reio}$ in Table 3 \& 4 of \citet[]{Paoletti:2021gzr} are consistent with our results, shown in Table \ref{tab:summary-Pk}. We believe the above consistency checks make the reconstruction technique a robust learning process and the inferences on the astrophysical parameters obtained therefrom are quite reliable that can be used for future analysis.

\section{Concluding Remarks \label{sec:conclusion}} 

In this article, we have investigated how GPR can enhance our understanding the history of reionization and associated astrophysical parameters during this period. We trained the GP algorithm for model-independent reconstruction of the UV luminosity function using UV17 data, which is then combined with the Planck optical depth and QHII Lyman-$\alpha$ measurements to undertake a non-parametric reconstruction of the reionization history. For a robust analysis, we allowed the parameters of the logarithmic double power law and kernel hyperparameters to vary freely during GP training, expanding upon previous literature \citep[]{Ishigaki_2018, Hazra:2019wdn, Adak:2024urf}, and found our results to be consistent. The clumping factor $C_{\rm HII}$ was kept free, rather than being fixed to specific values like 3 \citep[]{Hazra:2019wdn} or 5 \citep[]{Adak:2024urf}, allowing us to explore potential redshift variations and compare the roles of astrophysical parameters in shaping the reionization history.

We extended our analysis to explore previously unexplored applications of GPR in reionization. This involved separately considering the global 21-cm signal and the 21-cm power spectrum to investigate their roles in learning the astrophysical parameters and reionization history. For the global 21-cm signal, we trained GPR using SARAS mock data and found that its inclusion in the reconstruction pipeline significantly improves the bounds on the astrophysical parameters. We presented the 21-cm power spectrum analysis based on the reconstructed reionization history, by making modifications to the Boltzmann solver \texttt{CLASS} and MCMC code \texttt{MontePython}. Our findings indicate that the power spectrum analysis for the modified reionization history with future SKA will help improve the bounds on six cosmological parameters and two astrophysical parameters. Additionally, both of the astrophysical parameters prefer values closer to the lower bounds of the baseline tanh reionization model.

For the {SARAS-inclusive analysis}, we rely on mock global signal data generated using the best-fit parameters from the SARAS and SKA-excluded runs (i.e., UV17+QHII+Planck). While the posterior in the SARAS-excluded run favored higher values of $C_{\rm HII} \sim 3-4$, the SARAS mock itself was generated with a \emph{fixed} $C_{\rm HII} \sim 4 - 5$ in this range, along with the corresponding best-fit values for other parameter, such as $\log_{10}\langle f_{\rm esc}\,\xi_{\rm ion}\rangle \sim 24.1 -24.4$, respectively. Consequently, in the SARAS-inclusive analysis Fig. \ref{fig:saras}(a), the posterior naturally peaks tightly around $C_{\rm HII} \sim 3-4$, reflecting the mock input bias rather than a rejection of lower values from real data. In the {SKA-based analysis}, mock power spectra are generated within the \texttt{MontePython} pipeline using the Planck 2018 cosmology and the GP-reconstructed reionization history (Table~1). Although the reconstruction accounts for the redshift evolution of the ionization fraction, the astrophysical parameters (e.g., $\log_{10} \langle f_{\rm esc}, \xi_{\rm ion} \rangle$ and $C_{\rm HII}$) are effectively averaged over redshift. From the power spectrum analysis with Planck+SKA+Ly$\alpha$+UV17 data we found the 1$\sigma$ bound on $C_{\rm HII} \sim 1$. 

Previous studies discuss the possiblity of a wide range of $C_{\rm HII}$ values: hydrodynamical simulations of the early universe suggested $C_{\rm HII} \sim 30$ \citep[e.g.][]{Gnedin:1996qr}, implying that recombination played a significant role in the progression of reionization. Later dark-matter-only simulations yielded $C_{\rm HII} \sim 10$ \citep[]{Iliev:2004ak, Iliev:2006sw}, while radiation-hydrodynamic studies indicate lower values of $2$–$3$ \citep[]{McQuinn_2011}, though recent work points to somewhat higher estimates \citep[e.g.][]{Kannan:2021xoz}. A redshift-dependent model, $C_{\rm HII} = 2.9 \times \left(\tfrac{1+z}{2}\right)^{-1.1}$ \citep[]{Paoletti:2024lji, 2011arXiv1108.3334S}, suggests that low values (near unity) inferred from power spectrum–based mocks are consistent with early reionization, while higher values from global-signal mocks reflect later stages.

Again, while $f_{\rm esc}$ is particularly difficult to constrain at $z \sim 2$–$4.5$ \citep{Inoue:2014zna, Robertson:2021ljt}, early studies favored low values ($f_{\rm esc} \approx 0.2$) consistent with CMB and HFF data \citep{Robertson:2013bq}. More recent JWST results indicate higher $\xi_{\rm ion}$ at $z > 9$, highlighting a degeneracy between $\xi_{\rm ion}$ and $f_{\rm esc}$ \citep{Munoz:2024fas, Simmonds_2024, Atek_nature}. Studies also suggest $f_{\rm esc}$ evolves with redshift \citep{Kulkarni_2019, Finkelstein_2019, Cain_2021, Katz:2022usl}, though a constant $f_{\rm esc} \sim 0.06$–$0.1$ for $z \geq 6$ is consistent with current data \citep{Mitra:2023yyv}. Within $\Lambda$CDM, the product $\xi_{\rm ion} f_{\rm esc}$ varies widely across reionization models, complicating cosmological interpretations \citep{Hazra:2019wdn, Paoletti:2021gzr, Chatterjee:2021ygm, Dey:2022ini, Dey:2023sxx, Paoletti:2024lji}. Our results for $\log_{10} \langle f_{\rm esc} \, \xi_{\rm ion} \rangle$ and $z_{\rm reio}$ agree well with existing constraints.

However, it is important to emphasize that the constraints derived from SARAS and SKA mock analyses are dependent on underlying mock datasets.  To investigate this difference in the resulting $C_{\rm HII}$ constraints between the SARAS- vs SKA-inclusive analyses, we further regenerated the SARAS mock with $C_{\rm HII} \sim 1$ and a lower input value of $\log_{10}\langle f_{\rm esc}\,\xi_{\rm ion}\rangle \sim 23.2 - 23.4$. In this case, we recover $C_{\rm HII} \sim 1$ as an allowed value, but observe a significant shift in $\log_{10}\langle f_{\rm esc}\,\xi_{\rm ion}\rangle$, This again confirms a strong degeneracy between $C_{\rm HII}$ and $\log_{10}\langle f_{\rm esc} \, \xi_{\rm ion}\rangle$. Therefore, when fitting the mock SARAS data, the posterior for $C_{\rm HII}$ peaks sharply around the injected value not due to rejection of other values from actual data, but because the mock input enforces this value, maintaining the corelation with $\log_{10}\langle f_{\rm esc}\,\xi_{\rm ion}\rangle$.

As a result, the apparent variation in $C_{\rm HII}$ and $\log_{10} \langle f_{\rm esc} \, \xi_{\rm ion} \rangle$ across SARAS- and SKA-inclusive analyses primarily reflects their differing redshift sensitivities. Any true underlying redshift evolution in $C_{\rm HII}(z)$ or $\log_{10} \langle f_{\rm esc} \, \xi_{\rm ion} \rangle$, which is not explicitly modeled in this work, would naturally lead to different recovered values in the SARAS and SKA analyses pipelines, without implying a contradiction. Furthermore, we would like to clarify that the recovered parameter values are reflective of different effective redshift regimes, and should not be compared without accounting for potential redshift evolution. This underscores our main conclusion: both global signal (e.g. SARAS) and power spectrum (e.g. SKA) probes offer complementary insight into EoR astrophysics, and future datasets will be key in breaking degeneracies between the astrophysical parameters.

Thus, differences in $C_{\rm HII}$ from SKA- vs. SARAS-based analyses reflect redshift evolution rather than inconsistency. Our goal here is not to assert statistical consistency or tension between the values of $C_{\rm HII}$ across analyses, but rather to emphasize the complementary nature of the information the global signal or power spectrum can provide. The global signal and power spectrum analyses are two important and distinct types of observables in 21 cm cosmology. In global signal analysis, only the spatially averaged (global) quantities are considered, without accounting for perturbations. In contrast, power spectrum analysis explicitly includes the spatial fluctuations of different components. As a result, these two analyses probe different aspects of the 21 cm signal and are fundamentally different in nature. Thus in this paper two analysis gives the complete pictures of the reionzation history of the Universe. As our analysis is based on idealized mock datasets, we refrain from making strong statistical claims at this stage. 

Thus, our analysis demonstrates that GPR-based reconstruction performs well in the context of reionization, both for the global 21-cm signal and the 21-cm power spectrum, in combination with other relevant data sets.  The results derived from our method are hence reliable and can be used in future reionization studies. That said, our reconstruction framework can be extended in several directions. The current power spectrum analysis is based on Eq. \eqref{eq:P21} and does not yet include the redshift evolution of $\log_{10} \langle f_{\rm esc} \, \xi_{\rm ion} \rangle$, though we do incorporate redshift evolution in $C_{\rm HII}$. A more rigorous analysis will involve reionization simulation for estimating the actual bounds on the astrophysical parameters and possible reflections on SKA \citep[]{Mangena:2020jdo}. Further, our analysis is based only on the $\Lambda$CDM model; as our future goal, we will explore beyond-$\Lambda$CDM models and investigate their impact on the reionization period.

For reconstruction purposes, the kinetic Sunyaev-Zeldovich data \citep[]{Jain:2023jpy} can also be added on top of current data sets and the possible consequences can be investigated. Lastly, while we have utilized the GPR technique as an ML tool for learning, future investigations will explore various ML techniques \citep[]{Sohn:2022jsm, Gomez-Vargas:2022bsm, Shah:2024slr, Mukherjee:2024akt} for training and testing in the context of reionization. We look forward to pursuing these avenues in future work.

\software{} \texttt{emcee} \citep[]{Foreman-Mackey:2012any}, \texttt{CLASS} \citep[]{Blas_2011}, \texttt{MontePython} \citep[]{Audren:2012wb, Brinckmann:2018cvx}, \texttt{GetDist} \citep[]{Lewis:2019xzd}

\begin{acknowledgments}
We acknowledge Rahul Shah and Debarun Paul for their useful comments. PM would like to thank Dhiraj Kumar Hazra and Anjan Ananda Sen for discussions. PM acknowledges funding from the {Anusandhan National Research Foundation (ANRF), Govt of India} under the National Post-Doctoral Fellowship (File no. PDF/2023/001986). AD thanks IOP, Bhubaneswar for financial support. SP thanks the Department of Science and Technology, Govt. of India for partial support through Grant No. NMICPS/006/MD/2020-21 {and also the ANRF, Govt. of India for partial support through Project No. CRG/2023/003984. The authors gratefully acknowledge the computational facility of ISI Kolkata and the use of the High-Performance Computing facility, Pegasus, at IUCAA, Pune, India.}
\end{acknowledgments}

\bibliography{references}{}
\bibliographystyle{aasjournal}

\end{document}